\input harvmac
\input epsf.tex
\def\N{{\cal N}}

\def\lfm#1{\medskip\noindent\item{#1}}

\noblackbox
%
%

\batchmode
  \font\bbbfont=msbm10
\errorstopmode
\newif\ifamsf\amsftrue
\ifx\bbbfont\nullfont
  \amsffalse
\fi
\ifamsf
\def\IR{\hbox{\bbbfont R}}
\def\IZ{\hbox{\bbbfont Z}}
\def\IF{\hbox{\bbbfont F}}
\def\IP{\hbox{\bbbfont P}}
\else
\def\IR{\relax{\rm I\kern-.18em R}}
\def\IZ{\relax\ifmmode\hbox{Z\kern-.4em Z}\else{Z\kern-.4em Z}\fi}
\def\IF{\relax{\rm I\kern-.18em F}}
\def\IP{\relax{\rm I\kern-.18em P}}
\fi

\def\ev#1{\langle#1\rangle}
\

\lref\IW{
K.~Intriligator and B.~Wecht,
``The exact superconformal R-symmetry maximizes a,''
Nucl.\ Phys.\ B {\bf 667}, 183 (2003)
[arXiv:hep-th/0304128].
}
\lref\IWbar{
K.~Intriligator and B.~Wecht,
``Baryon charges in 4D superconformal field theories and their AdS  duals,''
Commun.\ Math.\ Phys.\  {\bf 245}, 407 (2004)
[arXiv:hep-th/0305046].
}
\lref\Tachikawa{
Y.~Tachikawa,
  ``Five-dimensional supergravity dual of a-maximization,''
  arXiv:hep-th/0507057.
}
\lref\JEHO{
  J.~Erdmenger and H.~Osborn,
  ``Conserved currents and the energy-momentum tensor in conformally  invariant
  theories for general dimensions,''
  Nucl.\ Phys.\ B {\bf 483}, 431 (1997)
  [arXiv:hep-th/9605009].
}
\lref\ButtiSW{
  A.~Butti, D.~Forcella and A.~Zaffaroni,
  ``The dual superconformal theory for L(p,q,r) manifolds,''
  arXiv:hep-th/0505220.
}
\lref\Tseytlin{
  F.~Bastianelli, S.~Frolov and A.~A.~Tseytlin,
  ``Three-point correlators of stress tensors in maximally-supersymmetric
  conformal theories in d = 3 and d = 6,''
  Nucl.\ Phys.\ B {\bf 578}, 139 (2000)
  [arXiv:hep-th/9911135].
}
\lref\BHac{
  S.~Benvenuti and A.~Hanany,
  ``New results on superconformal quivers,''
  arXiv:hep-th/0411262.
}
\lref\Anselmi{
  D.~Anselmi,
  ``Central functions and their physical implications,''
  JHEP {\bf 9805}, 005 (1998)
  [arXiv:hep-th/9702056].
}
\lref\ISmir{
  K.~A.~Intriligator and N.~Seiberg,
  ``Mirror symmetry in three dimensional gauge theories,''
  Phys.\ Lett.\ B {\bf 387}, 513 (1996)
  [arXiv:hep-th/9607207].
}
\lref\ISmirr{
  O.~Aharony, A.~Hanany, K.~A.~Intriligator, N.~Seiberg and M.~J.~Strassler,
  ``Aspects of N = 2 supersymmetric gauge theories in three dimensions,''
  Nucl.\ Phys.\ B {\bf 499}, 67 (1997)
  [arXiv:hep-th/9703110].
}
\lref\Witten{
  E.~Witten,
  ``Anti-de Sitter space and holography,''
  Adv.\ Theor.\ Math.\ Phys.\  {\bf 2}, 253 (1998)
  [arXiv:hep-th/9802150].
}
\lref\Maldacena{
  J.~M.~Maldacena,
  ``The large N limit of superconformal field theories and supergravity,''
  Adv.\ Theor.\ Math.\ Phys.\  {\bf 2}, 231 (1998)
  [Int.\ J.\ Theor.\ Phys.\  {\bf 38}, 1113 (1999)]
  [arXiv:hep-th/9711200].
}
\lref\PKFL{
  P.~Kraus and F.~Larsen,
  ``Attractors and black rings,''
  arXiv:hep-th/0503219.
}
\lref\GubserVD{
  S.~S.~Gubser,
  ``Einstein manifolds and conformal field theories,''
  Phys.\ Rev.\ D {\bf 59}, 025006 (1999)
  [arXiv:hep-th/9807164].
}
\lref\GKP{
  S.~S.~Gubser, I.~R.~Klebanov and A.~M.~Polyakov,
  ``Gauge theory correlators from non-critical string theory,''
  Phys.\ Lett.\ B {\bf 428}, 105 (1998)
  [arXiv:hep-th/9802109].
}
\lref\GMSW{
  J.~P.~Gauntlett, D.~Martelli, J.~F.~Sparks and D.~Waldram,
  ``A new infinite class of Sasaki-Einstein manifolds,''
  arXiv:hep-th/0403038.
}
\lref\GatesNR{
  S.~J.~Gates, M.~T.~Grisaru, M.~Rocek and W.~Siegel,
  ``Superspace, Or One Thousand And One Lessons In Supersymmetry,''
  Front.\ Phys.\  {\bf 58}, 1 (1983)
  [arXiv:hep-th/0108200].
}
\lref\AnselmiXK{
  D.~Anselmi,
  ``Anomalies, unitarity, and quantum irreversibility,''
  Annals Phys.\  {\bf 276}, 361 (1999)
  [arXiv:hep-th/9903059].
}
\lref\FW{
  S.~Franco, A.~Hanany, D.~Martelli, J.~Sparks, D.~Vegh and B.~Wecht,
  ``Gauge theories from toric geometry and brane tilings,''
  arXiv:hep-th/0505211.
}
\lref\HEK{
  C.~P.~Herzog, Q.~J.~Ejaz and I.~R.~Klebanov,
 ``Cascading RG flows from new Sasaki-Einstein manifolds,''
  JHEP {\bf 0502}, 009 (2005)
  [arXiv:hep-th/0412193].
}
\lref\FMMR{
  D.~Z.~Freedman, S.~D.~Mathur, A.~Matusis and L.~Rastelli,
  ``Correlation functions in the CFT($d$)/AdS($d+1$) correspondence,''
  Nucl.\ Phys.\ B {\bf 546}, 96 (1999)
  [arXiv:hep-th/9804058].
}
\lref\Fabbri{
  D.~Fabbri, P.~Fre', L.~Gualtieri, C.~Reina, A.~Tomasiello, A.~Zaffaroni and A.~Zampa,
  ``3D superconformal theories from Sasakian seven-manifolds: New  nontrivial
  evidences for AdS(4)/CFT(3),''
  Nucl.\ Phys.\ B {\bf 577}, 547 (2000)
  [arXiv:hep-th/9907219].
}

\lref\MHKS{
  M.~Henningson and K.~Skenderis,
  ``The holographic Weyl anomaly,''
  JHEP {\bf 9807}, 023 (1998)
  [arXiv:hep-th/9806087].
}
\lref\HOsc{
  H.~Osborn,
  ``N = 1 superconformal symmetry in four-dimensional quantum field theory,''
  Annals Phys.\  {\bf 272}, 243 (1999)
  [arXiv:hep-th/9808041].
}
\lref\WeinbergKK{
  S.~Weinberg,
  ``Charges From Extra Dimensions,''
  Phys.\ Lett.\ B {\bf 125}, 265 (1983).
}
\lref\MS{
  D.~Martelli and J.~Sparks,
  ``Toric geometry, Sasaki-Einstein manifolds and a new infinite class of
  AdS/CFT duals,''
  arXiv:hep-th/0411238.
}

\lref\BBC{
  M.~Bertolini, F.~Bigazzi and A.~L.~Cotrone,
  ``New checks and subtleties for AdS/CFT and a-maximization,''
  JHEP {\bf 0412}, 024 (2004)
  [arXiv:hep-th/0411249].
}
\lref\BHK{
  D.~Berenstein, C.~P.~Herzog and I.~R.~Klebanov,
  ``Baryon spectra and AdS/CFT correspondence,''
  JHEP {\bf 0206}, 047 (2002)
  [arXiv:hep-th/0202150].
}
\lref\HM{
  C.~P.~Herzog and J.~McKernan,
  ``Dibaryon spectroscopy,''
  JHEP {\bf 0308}, 054 (2003)
  [arXiv:hep-th/0305048].
}
\lref\GLMW{
  J.~P.~Gauntlett, S.~Lee, T.~Mateos and D.~Waldram,
  ``Marginal deformations of field theories with AdS(4) duals,''
  arXiv:hep-th/0505207.
}
\lref\HW{
  C.~P.~Herzog and J.~Walcher,
  ``Dibaryons from exceptional collections,''
  JHEP {\bf 0309}, 060 (2003)
  [arXiv:hep-th/0306298].
}
\lref\GKP{
  S.~S.~Gubser, I.~R.~Klebanov and A.~M.~Polyakov,
  ``Gauge theory correlators from non-critical string theory,''
  Phys.\ Lett.\ B {\bf 428}, 105 (1998)
  [arXiv:hep-th/9802109].
}
\lref\malda{
  J.~M.~Maldacena,
  ``The large N limit of superconformal field theories and supergravity,''
  Adv.\ Theor.\ Math.\ Phys.\  {\bf 2}, 231 (1998)
  [Int.\ J.\ Theor.\ Phys.\  {\bf 38}, 1113 (1999)]
  [arXiv:hep-th/9711200].
}
\lref\GSTii{
  M.~Gunaydin, G.~Sierra and P.~K.~Townsend,
  ``Gauging The D = 5 Maxwell-Einstein Supergravity Theories: More On Jordan
  Algebras,''
  Nucl.\ Phys.\ B {\bf 253}, 573 (1985).
}
\lref\Herzog{
  C.~P.~Herzog,
  ``Exceptional collections and del Pezzo gauge theories,''
  JHEP {\bf 0404}, 069 (2004)
  [arXiv:hep-th/0310262].
}
\lref\HKO{
  C.~P.~Herzog, I.~R.~Klebanov and P.~Ouyang,
  ``D-branes on the conifold and N = 1 gauge / gravity dualities,''
  arXiv:hep-th/0205100.
}

\lref\GK{
  S.~S.~Gubser and I.~R.~Klebanov,
  ``Baryons and domain walls in an N = 1 superconformal gauge theory,''
  Phys.\ Rev.\ D {\bf 58}, 125025 (1998)
  [arXiv:hep-th/9808075].
}
\lref\MSY{
  D.~Martelli, J.~Sparks and S.~T.~Yau,
  ``The geometric dual of a-maximisation for toric Sasaki-Einstein manifolds,''
  arXiv:hep-th/0503183.
}
\lref\DuffCC{
  M.~J.~Duff, C.~N.~Pope and N.~P.~Warner,
  ``Cosmological And Coupling Constants In Kaluza-Klein Supergravity,''
  Phys.\ Lett.\ B {\bf 130}, 254 (1983).
}
\lref\GRW{
  M.~Gunaydin, L.~J.~Romans and N.~P.~Warner,
  ``Compact And Noncompact Gauged Supergravity Theories In Five-Dimensions,''
  Nucl.\ Phys.\ B {\bf 272}, 598 (1986).
}
\lref\BF{
  S.~Benvenuti, S.~Franco, A.~Hanany, D.~Martelli and J.~Sparks,
  ``An infinite family of superconformal quiver gauge theories with
  Sasaki-Einstein duals,''
  arXiv:hep-th/0411264.
}
\lref\taumin{E. Barnes, E. Gorbatov, K. Intriligator, M. Sudano, and J. Wright, ``The exact
superconformal R-symmetry minimizes $\tau _{RR}$."}
 \lref\AEFJ{D.~Anselmi, J.~Erlich, D.~Z.~Freedman and A.~A.~Johansen,
``Positivity constraints on anomalies in supersymmetric gauge
theories,''
Phys.\ Rev.\ D {\bf 57}, 7570 (1998)
[arXiv:hep-th/9711035].
}
\lref\GST{
  M.~Gunaydin, G.~Sierra and P.~K.~Townsend,
  ``The Geometry Of N=2 Maxwell-Einstein Supergravity And Jordan Algebras,''
  Nucl.\ Phys.\ B {\bf 242}, 244 (1984).
}
\lref\AFGJ{D.~Anselmi, D.~Z.~Freedman, M.~T.~Grisaru and A.~A.~Johansen,
``Nonperturbative formulas for central functions of supersymmetric gauge
theories,''
Nucl.\ Phys.\ B {\bf 526}, 543 (1998)
[arXiv:hep-th/9708042].
}
\lref\deWitEQ{
  B.~de Wit and H.~Nicolai,
  ``N=8 Supergravity With Local SO(8) X SU(8) Invariance,''
  Phys.\ Lett.\ B {\bf 108}, 285 (1982).
}
\lref\HOAP{
  H.~Osborn and A.~C.~Petkou,
  ``Implications of conformal invariance in field theories for general
  dimensions,''
  Annals Phys.\  {\bf 231}, 311 (1994)
  [arXiv:hep-th/9307010].
}
\lref\FMMR{
  D.~Z.~Freedman, S.~D.~Mathur, A.~Matusis and L.~Rastelli,
  ``Correlation functions in the CFT($d$)/AdS($d+1$) correspondence,''
  Nucl.\ Phys.\ B {\bf 546}, 96 (1999)
  [arXiv:hep-th/9804058].
}
\lref\KBMC{
  K.~Behrndt and M.~Cvetic,
  ``Anti-de Sitter vacua of gauged supergravities with 8 supercharges,''
  Phys.\ Rev.\ D {\bf 61}, 101901 (2000)
  [arXiv:hep-th/0001159].
}
\lref\ButtiVN{
  A.~Butti and A.~Zaffaroni,
  ``R-charges from toric diagrams and the equivalence of a-maximization and
  Z-minimization,''
  arXiv:hep-th/0506232.
}
\def\drawbox#1#2{\hrule height#2pt
             \hbox{\vrule width#2pt height#1pt \kern#1pt \vrule
width#2pt}
                   \hrule height#2pt}

\def\Fund#1#2{\vcenter{\vbox{\drawbox{#1}{#2}}}}
\def\Asym#1#2{\vcenter{\vbox{\drawbox{#1}{#2}
                   \kern-#2pt       
                   \drawbox{#1}{#2}}}}

\def\sym{\Fund{6.5}{0.4} \kern-.5pt \Fund{6.5}{0.4}}
\Title{\vbox{\baselineskip12pt\hbox{hep-th/0507146}
\hbox{UCSD-PTH-05-10}}}
{\vbox{\centerline{Current Correlators and AdS/CFT Geometry}}}
\centerline{ Edwin Barnes, Elie Gorbatov, Ken Intriligator, and Jason Wright}
\bigskip
\centerline{Department of Physics} \centerline{University of
California, San Diego} \centerline{La Jolla, CA 92093-0354, USA}

\bigskip
\noindent
We consider current-current correlators in 4d $\N =1$ SCFTs, and also 3d
$\N =2$ SCFTs, in connection with AdS/CFT geometry.  The superconformal
$U(1)_R$ symmetry of the SCFT has the distinguishing property that, among
all possibilities, it minimizes the coefficient, $\tau _{RR}$ of its two-point
function.  We show that 
 the geometric Z-minimization condition of Martelli, Sparks, and Yau precisely implements
$\tau _{RR}$ minimization.    This gives a physical  proof that Z-minimization in geometry indeed correctly determines the superconformal R-charges of the field theory dual.   We further discuss and compare current two point functions
in field theory and  AdS/CFT and the geometry of Sasaki-Einstein manifolds.  
Our analysis gives new quantitative checks of the AdS/CFT correspondence.

\Date{July 2005}
\newsec{Introduction}

This work is devoted to the geometry / gauge theory interrelations of the AdS/CFT correspondence  \refs{\malda, \GKP, 
\Witten}, which has been much developed and checked over the past year (a
sample of recent references is \refs{\GMSW, \MS, \BBC, \BF, \HEK, \MSY, \FW, \ButtiSW}).

In the AdS/CFT correspondence \refs{\malda, \GKP, \Witten}, global currents $J^\mu _I$ ($I$ labels  the various currents) of the d-dimensional CFT couple to gauge fields in the $AdS_{d+1}$ bulk. The current two-point functions of the CFT are of fixed form, 
\eqn\jjev{\ev{J^\mu _I(x)J^\nu _J(y)}={\tau _{IJ}\over (2\pi )^d}(\partial ^2\delta _{\mu \nu}-\partial _\mu \partial _\nu){1\over (x-y)^{2(d-2)}},}
with only the coefficients $\tau _{IJ}$ depending on the theory and its dynamics.  Unitarity restricts $\tau _{IJ}$ to
be a positive matrix  (positive eigenvalues).   
The  coefficients $\tau _{IJ}$ map to the coupling constants of the corresponding gauge fields in $AdS_{d+1}$: writing
their kinetic terms as 
\eqn\adskt{S_{AdS_{d+1}}=\int d^dzdz_0\sqrt{g}\left[-{1\over 4}g^{-2}_{IJ}F^I_{\mu \nu}F^{\mu \nu J}+\dots\right],}
the relation is \FMMR:
\eqn\tauads{\tau _{IJ}={2^{d-2}\pi ^{d\over 2}\Gamma [d]\over (d-1)\Gamma[{d\over 2}]}L^{d-3}g_{IJ}^{-2},}
where $L$ is the $AdS_{d+1}$ length scale.   Our main interest here will be in the quantities
$\tau _{IJ}$, and comparing field theory results with the $AdS$ relation \tauads.  

We will here consider 4d $\N =1$ superconformal field theories, 3d $\N =2$ SCFTs, and their AdS duals, 
coming, respectively, from IIB string theory on $AdS_5\times Y_5$, 11d SUGRA or M-theory on $AdS_4\times Y_7$.  Supersymmetry requires $Y_5$ and $Y_7$ to be Sasaki-Einstein.   In general,
a Sasaki-Einstein space $Y_{2n-1}$ is the horizon of a 
non-compact local Calabi-Yau n-fold $X_{2n}=C(Y_{2n-1})$, with  conical metric 
\eqn\xandh{ds^2(C(Y_{2n-1}))=dr^2+r^2ds^2(Y_{2n-1}).} 
The gauge theories come from $N$ $D3$ or $M2$ branes at the tip of the cone. In
the large $N$ dual, the radial $r$ becomes that of $AdS_{d+1}$.  The dual to 4d
$\N=1$ SCFTs is IIB on
\eqn\iibadsy{AdS_5\times Y_5:\qquad ds_{10}^2={r^2\over L^2}\eta _{\mu \nu}dx^\mu dx^\nu +{L^2\over r^2}dr^2+L^2 ds^2(Y_5),}
and the dual to 3d $\N =2$ SCFTs is 11d SUGRA or M-theory with metric background 
\eqn\madsy{AdS_4\times Y_7:\qquad ds_{11}^2={r^2\over L^2}\eta _{\mu \nu}dx^\mu dx^\nu
+{L^2\over r^2}dr^2+(2L)^2ds^2(Y_7).}

The SCFTs  have a conserved, superconformal $U(1)_R$ current,  in the same supermultiplet as the stress tensor.  The scaling dimensions of chiral operators are related to their superconformal $U(1)_R$ charges by
\eqn\delR{\Delta = {d-1\over 2}R.}
There are also typically various non-R flavor currents, whose charges we'll write as $F_i$, with $i$ labeling the flavor symmetries. The 
superconformal $U(1)_R$ of RG fixed point SCFTs is then not determined by the symmetries alone, as the R-symmetry can mix with the flavor symmetries.  Some additional dynamical information is then needed to determine precisely which, among all possible R-symmetries, is the superconformal  one, in the stress tensor supermultiplet.  

On the field theory side, we presented a new condition in \taumin, which, 
in principle, uniquely determines the superconformal $U(1)_R$: among all possible trial R-symmetries, 
\eqn\rgen{R_t=R_0+\sum _i s_i F_i,}
the superconformal one is that which {\it minimizes} the coefficient $\tau _{R_tR_t}$ of its two point function \jjev.  An equivalent way to state this is that the two-point function of the superconformal R-current with all non-R flavor symmetries necessarily vanishes:
\eqn\taufrs{\tau _{Ri}=0 \qquad\hbox{for all non-R symmetries $F_i$}.}
(Our notation will always be that capital $I$ runs over all symmetries, including the 
superconformal $U(1)_R$, and lower case $i$ runs over the non-R flavor symmetries.)
We refer to the field theory condition of \taumin\ as ``$\tau _{RR}$ minimization".  The 
minimal value of $\tau _{R_tR_t}$ is then the coefficient, $\tau _{RR}$, of the
superconformal $U(1)_R$ current two-point function, which is related by supersymmetry
to the coefficient of the stress-tensor two-point function, 
\eqn\taurrc{\tau _{RR}\propto C_T.}

For the case of 4d $\N =1$ SCFTs, a-maximization \IW\ gives another way, besides $\tau _{RR}$ minimization, to determine the superconformal $U(1)_R$:
 the exact superconformal R-symmetry is that which 
(locally) maximizes  the combination of 't Hooft anomalies
\eqn\amax{a_{trial}(R_t)={3\over 32}(3\Tr R ^3-\Tr R).}
Equivalently, the superconformal $U(1)_R$ satisfies the 't Hooft anomaly identity \IW
\eqn\amaxi{9\Tr R^2F_i=\Tr F_i \qquad\hbox{for all flavor symmetries $F_i$.}}
a-maximization does not apply for 3d SCFTs, as there are there no 't Hooft anomalies. 

 The global symmetries of the $SCFT_d$ map
to the following gauge symmetries in the $AdS_{d+1}$ bulk:
\lfm{1.} The graviphoton, which maps to the superconformal $U(1)_R$, is a Kaluza-Klein gauge field, associated with the ``Reeb" Killing vector isometry of Sasaki-Einstein $Y_{2n-1}$.  The R-charge is normalized so that superpotential terms, which are related to the holomorphic $n$ form of $X_{2n}$, have charge $R=2$.
\lfm{2.} Any other Kaluza-Klein gauge fields, from any additional isometries of $Y_{2n-1}$.
These can be taken to be non-R symmetries, by taking the holomorphic n-form to be
neutral.  We refer to these as ``mesonic, non-R, flavor symmetries," because mesonic operators (gauge invariants not requiring an epsilon tensor) of the dual gauge theory can be charged under them.  When $Y_{2n-1}$ is toric, there is always (at least) a $U(1)^{n-1}$ group of mesonic, non-R flavor symmetries. 
\lfm{3.} Baryonic $U(1)^{b_*}$ gauge fields, from reducing Ramond-Ramond gauge fields on non-trivial 
cycles of $Y_{2n-1}$.  In particular, for IIB on $AdS_5\times Y_5$, there are $U(1)^{b_3}$ baryonic gauge fields come from reducing $C_4$ on the $b_3=$dim($H_3(Y_5)$) non-trivial 3-cycles of $Y_5$.  These are also non-R symmetries.  Baryonic $U(1)$ symmetries have the distinguishing property in the gauge theory that only baryonic operators, formed with
an epsilon tensor, are charged under them.  It was pointed out in \IWbar\ that 4d baryonic symmetries
have another distinguishing property: their cubic 't Hooft anomalies all vanish, $\Tr U(1)_B^3=0$, 
as seen from the fact that it's not possible to get the needed Chern-Simons term \Witten\ $A_B\wedge dA_B\wedge dA_B$ from reducing 10d string theory on $Y_5$.

In field theory, the superconformal $U(1)_R$ can, and generally does mix with the 
mesonic and baryonic 
\foot{A point of possible confusion:  as pointed out in \IW, the 
superconformal $U(1)_R$ does not mix with those baryonic symmetries which transform under
charge conjugation symmetry.  But the superconformal gauge theories associated with general $Y_{2n-1}$ are chiral, with no charge conjugation symmetries.  So the superconformal $U(1)_R$
can mix with these baryonic $U(1)$'s.}
flavor symmetries.   The correct superconformal $U(1)_R$ can, in principle,  be determined by $\tau _{RR}$ minimization  \taumin. 
$\tau _{RR}$ minimization is not especially practical to implement in field theory, because the coefficients \taufrs\ get quantum corrections.  But, on the AdS dual side, 
$\tau _{RR}$ minimization  becomes more useful and tractable, because the AdS duality
gives a weakly coupled dual description of $\tau _{R_0i}$ and $\tau_{ij}$, via \tauads.

The problem of determining the superconformal $U(1)_R$ in the field theory maps to a corresponding problem in the geometry: determining which $U(1)$, out of the $U(1)^n$ geometric
isometries of toric Sasaki-Einstein spaces, is that of the Reeb vector.  A solution of this mathematical problem was recently found by Martelli, Sparks, and Yau \MSY: the correct Reeb 
vector is that which minimizes the Einstein-Hilbert action on $Y_{2n-1}$ -- this is referred to as
``Z-minimization," \MSY.  The mathematical result of
\MSY\ was shown, on a case-by-case basis, to always lead to the same superconformal R-charges as found from a-maximization \IW\ in the corresponding field theory, but there was no general
proof as to why Z-minimization in geometry implements a-maximization in field theory.  In addition, Z-minimization applies to
general $Y_{2n-1}$, whereas a-maximization is limited to 4d SCFTs, and hence the case of
$AdS_5\times Y_{5}$.

Our main result will be to show that the Z-minimization of  Martelli, Sparks, and Yau \MSY\
is precisely equivalent to ensuring that the $\tau _{RR}$ minimization conditions \taufrs\ 
of \taumin\ are satisifed, i.e. {\it Z-minimization $= \tau _{RR}$ minimization}.
This demonstrates that Z-minimization in the geometry indeed determines the correct superconformal R-symmetry of the dual SCFT, not only for 4d SCFTs, but also for
3d SCFTs with dual \madsy.   We will also explain why it's OK that the $U(1)^{b_*}$ baryonic $U(1)$ symmetries did not enter into the geometric Z-minimization  of \MSY: the condition \taufrs\ is
automatically satisfied in the string theory constructions for all baryonic symmetries.

The outline of this paper is as follows. In sect. 2, we review relations in 4d $\N=1$ field theory for the  current two-point functions, and the 't Hooft anomalies of the superconformal $U(1)_R$.  We then show that these relations are satisfied by the effective $AdS_5$ bulk SUGRA theory, thanks to the structure of real special geometry.  In particular, the kinetic
terms in the $AdS_5$ bulk are related to the Chern-Simons terms, which
yield the 't Hooft anomalies of the dual SCFT.  In the following sections, we discuss how
these kinetic terms are obtained from the geometry of $Y$; it would be interesting to 
also directly obtain the Chern-Simons terms from the geometry of $Y$, but that will not be 
done here.  In sect. 3, we discuss the contributions to the kinetic terms in the $AdS$ bulk.
As usual, Kaluza-Klein gauge fields get a contribution, with coefficient $(g_{ij}^{-2})^{KK}$,
{}from reducing the Einstein term in the action on $Y$.  Because of the background flux in $Y$, there is also a contribution $(g^{-2}_{IJ})^{CC}$ from reducing the Ramond-Ramond $C$ field kinetic terms on $Y$.  We point out (closely following \DuffCC)
that these two contributions always have the fixed ratio: $(g^{-2}_{IJ})^{CC}=\half (D_c-1)(g_{IJ}^{-2})^{KK}$, for any Einstein manifold
$Y$ of dimension $D_c$.  This relation will be used, and checked, in following sections. 
For the baryonic gauge fields, there is only the contribution $(g^{-2}_{IJ})^{CC}$, from
reducing the Ramond-Ramond kinetic term on $Y$.

In sect. 4, we discuss generally how the gauge fields $A_I$ alter Ramond-Ramond flux
background, and thereby alter the Ramond-Ramond field at linearized level, as $\delta C=\sum _I \omega _I\wedge A_I$, for some particular $2n-3$ forms $\omega _I$ on $Y$. 
We discuss how the $A_I$ charges of branes wrapped on supersymmetric cycles can be
obtained by integrating $\omega _I$ over the cycle, and how the Ramond-Ramond
contribution to the gauge kinetic terms is written as $\sim \int _Y\omega _I\wedge *\omega _J$.   In
sect. 5, we review some aspects of Sasaki-Einstein geometry, and the analysis of \BHK\ for
how to determine the form $\omega _R$ for the $U(1)_R$ gauge field. 
In sect. 6, we generalize this to determine the forms $\omega _I$ for the non-R isometry
and baryonic gauge fields.  In sect. 7, we give expressions for the gauge kinetic
terms $g^{-2}_{IJ}$, and thereby the current-current two-point function coefficients
$\tau _{IJ}$ that we are interested in, in terms of integrals $\sim \int _Y \omega _I\wedge *\omega _J$ of these forms.  We note that this immediately implies that there is never
any mixing in the kinetic terms between Kaluza-Klein isometry gauge fields and 
the baryonic gauge fields, i.e. that 
\eqn\kkbar{\tau _{IJ}=0 \quad\hbox{automatically, for $I=$ Kaluza-Klein and $J=$ baryonic}.}
This shows that our condition \taufrs\ for the $U(1)_R$ is automatically satisfied,
for all baryonic symmetries, by taking $U(1)_R$ to be purely a Kaluza-Klein isometry gauge
field, without any mixing with the baryonic symmetries.    For the mesonic, non-R isometry 
gauge fields, the condition \taufrs\ becomes
\eqn\hintzero{\int _Y g_{ab}K^aK_i^b vol(Y)=0,}
which give conditions to determine the $U(1)_R$ isometry Killing vector $K^a$.  The condition \hintzero\ must hold for every non-R isometry Killing vector of $Y$,
i.e. for every Killing vector $K_i^a$ under which the
the holomorphic $n$ form of $C(Y_{2n-1})$ is neutral. 

In sect. 8, we summarize the results of Martelli, Sparks, and Yau \MSY\ for toric $C(Y)$. 
Then $Y_{2n-1}$ always has at least $U(1)^n$ isometry, associated with shifts of
toric coordinates $\phi _i$, and the $U(1)_R$ Killing
Reeb vector $K^a$ is given by some components $b_i$, $i=1\dots n$,  in this basis.  The volume of $Y$
and its supersymmetric cycles are completely determined by the $b_i$, without needing
to know the metric on $Y$.  And the $b_i$ are themselves determined by Z-minimization \MSY, which is minimization of the Einstein-Hilbert action on $Y$.  In sect. 9, we point
out that Z-minimization is precisely equivalent to $\tau _{RR}$ minimization.  We also discuss
the flavor charges of wrapped branes.  In sect. 10, we illustrate our results for the $Y^{p,q}$ examples of
\refs{\GMSW, \MS}.   We find the forms $\omega _I$, and thereby use the flavor charges of wrapped branes.  We also compute from the geometry of $Y$ the gauge kinetic term 
coefficients, and thus  the current-current two-point function coefficients $\tau _{IJ}$.  
These quantities, computed from the geometry of $Y$, match with those computed
in the dual field theory of \BF; this gives new checks of the AdS/CFT correspondence for these theories.

 In the final stages of writing up this paper, the very interesting work \ButtiVN\ appeared, in which it was mathematically shown that the Z-function \MSY\ of 5d toric 
Sasaki-Einstein $Y_5$ and the $a_{trial}$ function \IW\ of the dual quiver 4d gauge theory
are related by $Z(x,y)=1/a(x,y)$ (even before extremizing).  The approach and results of our paper  are orthogonal and complementary to those of \ButtiVN.  Also in 
the final stages of writing up this paper, the work \Tachikawa\ appeared, which significantly overlaps with the approach of section 2 of our paper, and indeed goes further along those
lines than we did here.  

\newsec{4d $\N =1$ SCFTs and real special geometry}

This section is somewhat orthogonal to the rest of the paper.  The rest of this paper is
devoted to deriving the $AdS$ bulk gauge field kinetic terms $g^{-2}_{IJ}$ in 
\adskt\ and \tauads\ directly from the geometry of $Y$.  In the present section, without
explicitly considering $Y$, we will discuss how the various identities of 4d $\N =1$ SCFTs
are guaranteed to also show up in the effective $AdS_5$ SUGRA theory, thanks to the
structure of real, special geometry.

Because the superconformal R-current is in the same supermultiplet as the stress tensor, 
their two-point function coefficients are proportional, $\tau _{RR}\propto C_T$.  Also, in
4d $C_T\propto c$, with $c$ the conformal anomaly coefficient in 
\eqn\ca{\ev{T^\mu _\mu}={1\over 120}{1\over (4\pi )^2}\left (c(\hbox{Weyl})^2-{a\over 4}(\hbox{Euler})\right).}
So $\tau _{RR}\propto c$; more precisely, 
\eqn\taurrct{\tau _{RR}={16\over 3}c,}
with $c$ normalized such that $c=1/24$ for a free $\N =1$ chiral superfield.
Supersymmetry also relates $a$ and $c$ in \ca\ to 
the 't Hooft anomalies of the superconformal $U(1)_R$ \AEFJ:
\eqn\acthooft{a={3\over 32}(3\Tr R^3-\Tr R)\qquad c={1\over 32}(9\Tr R^3-5\Tr R).}
Combining \taurrct\ and \acthooft, we have
\eqn\susyreln{\tau _{RR}={3\over 2}\Tr R^3 -{5\over 6}\Tr R,}
The flavor current two-point functions are also given by 't Hooft anomalies \AEFJ:
\eqn\RFF{\tau _{ij}=-3\Tr RF_i F_j.}

There are precise  analogs to the above relations in the 
effective\foot{The 5d SUGRA theory suffices for studying current two-point functions,
and relations to 't Hooft anomalies, even if there is no full, consistent truncation from 10d
to an effective 5d  theory.}
  5d $\N =2$ bulk gauged U(1) supergravity; this is not surprising given that, on both
sides of the duality, these relations come from the same $SU(2,2|1)$ superconformal symmetry  group.  

The bosonic part of the effective 5d Lagrangian is \GST\ (also see e.g. \KBMC) ${\cal L}^{\rm{bosonic}}=$
\eqn\lbose{\sqrt{|g|}\left[{1\over 2}R-{1\over 2}G_{ij}\partial _\mu \phi ^i\partial ^\mu \phi ^j-{1\over 4}g^{-2}_{IJ}F_{\mu \nu}^IF^{\mu \nu J}-V(X)\right]+{1\over 48}C_{IJK}A^I\wedge F^J\wedge F^K}
where, to simplify expressions, we'll set the 5d gravitational constant
$\kappa _5=1$ in this section. There are $n_V+1$ gauge
fields, $I=1\dots n_V+1$, one of them being the graviphoton, which corresponds to the
superconformal $U(1)_R$ in the 4d SCFT.  The $n_V$ gauge fields correspond to
the non-R (i.e. the gravitino is neutral under them) flavor symmetries, which reside in current supermultiplets $J_i$, $i=1\dots n_V$; the first component
of this supermultiplet is a scalar, which couples to the scalars $\phi ^i$ in \lbose.  The 
scalars of the $n_V$ vector multiplets
are constrained by real special geometry to the space 
\eqn\cijkc{{{\cal N}}\equiv  {1 \over 6}C_{IJK}X^IX^JX^K=1.}

The kinetic terms are all determined by the Chern-Simons coefficients $C_{IJK}$.  In 
particular, the gauge field kinetic term coefficients $g^{-2}_{IJ}$ are given by 
\eqn\gisi{g^{-2}_{IJ}=-\half \partial _I\partial _J\ln {\cal N}|_{\N =1}=-\half (C_{IJK}X^K-X_IX_J),}
where $X_I\equiv \half C_{IJK}X^JX^K.$  
In a given vacuum, where $X^I$ has expectation values satisfying \cijkc, the $n_V$ scalars
in \lbose\ are given by the tangents $X^I_i$ to the surface \cijkc, which satisfy
\eqn\cijkct{C_{IJK}X^I_iX^JX^K=0.}
This can be written as $X_IX^I_i=0$. 
The vacuum expectation value $X^I$ picks out the direction of the graviphoton $A_R$,
and the tangents $X^I_i$ pick out the direction of the non-R flavor gauge fields:
\eqn\gfdict{A^I =\alpha X^I A_R+X^I_i A_i,}
with $\alpha$ a normalization factor, to ensure that the R-symmetry is properly normalized, to give the gravitinos charges $\pm 1$.  The correct value is $\alpha = 2L/3$,
where $L$ is the $AdS_5$ length scale, related to the value of the potential at its minimum by
$\Lambda = -6/L^2$.   

Using \gfdict\ and \gisi, we can compute the kinetic
term coefficients for the graviphoton and non-R gauge fields.  Using \tauads\ to convert
these into the current-current 2-point function coefficients, we have for the R-symmetry/graviphoton kinetic term
\eqn\taurrx{\tau _{RR}=8\pi ^2Lg_{RR}^{-2}=8\pi ^2L\alpha ^2g^{-2}_{IJ}X^IX^J=12\pi ^2 L\alpha ^2.}
For the $n_V$ non-R gauge fields, we have 
\eqn\tauijx{\tau _{ij}=8\pi ^2Lg^{-2}_{ij}=8\pi ^2Lg^{-2}_{IJ}X^I_iX^J_j=-4\pi ^2LC_{IJK}X_i^IX_j^JX^K.}
It also follows from \gisi\ and \cijkct, $X_IX^I_i=0$, that there is no kinetic term mixing between the graviphoton 
and the non-R gauge fields:
\eqn\tanr{\tau _{Ri}=8\pi ^2 Lg^{-2}_{Ri}=8\pi ^2L\alpha g^{-2}_{IJ}X^I_iX^J=0 \qquad \hbox{for all $i=1\dots n_V$.}}
This matches with the general SCFT field theory result \taufrs\ of \taumin. 

The Chern-Simons terms for the graviphoton and flavor gauge fields are similarly
found from \gfdict.  We'll normalize them as $C_{IJK}/48=k_{IJK}/96\pi ^2$, where $k_{IJK}$
is the properly normalized 5d Chern-Simons coefficients, which map \Witten\ to the
't Hooft anomalies of the gauge theory:
\eqn\krrr{\Tr R^3=k_{RRR}=2\pi ^2 \alpha ^3C_{IJK}X^I{X^J}X^K=12\pi ^2\alpha ^3,}
\eqn\krri{\Tr R^2F_i=k_{RRj}=2\pi ^2\alpha ^2C_{IJK}X^IX^JX^K_i=0,}
where we used \cijkct, and also
\eqn\krij{\Tr RF_iF_j=k_{Rij}=2\pi ^2\alpha C_{IJK}X^IX^J_iX^K_j.}

The field theories with (weakly coupled) AdS duals generally have $\Tr R=0$ and also
$\Tr F_i=0$.  The result \krri\ then reproduces the 't Hooft anomaly identity \amaxi\ of \IW. 
For $\Tr R=0$, \susyreln\ becomes $\tau _{RR}={3\over 2}\Tr R^3$, which is reproduced
by \taurrx\ and \krrr\ for
$\alpha = 2L/3$ in \gfdict.  Also the relation \acthooft\ of \AFGJ, which for $\Tr R=0$
is  $a=c={9\over 32}\Tr R^3$, is also reproduced by \krrr\ for $\alpha = 2L/3$, since
the result of \MHKS\ is  $a=c=L^3\pi ^2$ in
$\kappa _5=1$ units.  The relation \RFF\ is also reproduced, for $\alpha =2L/3$, by 
\tauijx\ and \krij. 

In later sections, we will be interested in computing the $AdS_5$ gauge field kinetic
terms $\tau _{IJ}$ directly from IIB string theory on $AdS_5\times Y_5$.
  To connect with the above expressions, we restore the factors
of $\kappa _5$ via dimensional analysis, and convert using
\eqn\kappavis{{L^3\over \kappa _5 ^2}={L^3\over 8\pi G_5}={L^8Vol(Y_5)\over 8\pi G_{10}}={N^2\over 4}{\pi \over Vol(Y_5)},}
where 
$Vol(Y_5)$ is the dimensionless volume of $Y_5$, with factors of its length scale, which coincides with the $AdS_5$ length scale $L$, factored out.   The last equality of \kappavis\ uses the flux quantization / brane tensions relation (see \HKO\ and references therein) 
\eqn\fluxq{2\sqrt{\pi}\kappa_{10}^{-1}L^4Vol(Y_5)={L^4Vol(Y_5)\over \sqrt{2G_{10}}}=N\pi .}
E.g. using \kappavis\ the result of \MHKS\ becomes
\GubserVD
\eqn\hsac{a=c={L^3\pi ^2\over \kappa _5 ^2}= {N^2\over 4}{\pi ^3\over Vol(Y_5)},}
and  \taurrx\ for $\alpha = 2L/3$ becomes 
\eqn\taurr{\tau _{RR}={16\pi ^2\over 3}{L^3\over \kappa _5 ^2}=
{4N^2\over 3}{\pi ^3\over Vol (Y_5)}.}

In the following sections, we will directly compute the $\tau _{IJ}$ kinetic terms from
reducing SUGRA on $Y$.  One could also directly determine the Chern-Simons
coefficients $C_{IJK}$ from reduction on $Y$, but doing so would require going
beyond our linearized analysis, and we will not do that here.  It would be nice to 
extend our analysis to compute the $C_{IJK}$ from $Y$, and explicitly verify that the special
geometry relations reviewed in the present section are indeed satisfied.

\newsec{Kaluza-Klein gauge couplings: a general relation for Einstein spaces}

Our starting point is the Einstein action in 
$D_t=D+D_c$ spacetime dimensions, along with the Ramond-Ramond gauge field 
kinetic terms:
\eqn\actfull{{1\over 16\pi G_{D_t}}\int \left(R_{D_t}*1 -
{1\over 4} F\wedge * F \right).}
We'll be interested in fluctuations of this action around a background solution of the form
$M_D\times Y$, with $M_D$ non-compact and $Y$ compact, of dimension $D_c\equiv p+2$, with flux
\eqn\fluxgen{F_{p+2}^{bkgd}=(p+1)m^{-(p+1)}vol(Y),}
and metric
\eqn\metricm{ds^2=ds^2_M+m^{-2}ds^2_Y.}
Here $m^{-1}$ is the length scale of $Y$, which we'll always factor out explicitly; $vol(Y)$
is the volume form of $Y$, with the length scale $m^{-1}$ again factored out. 
(We always use lower case $vol(Y)$ for a volume form, and upper case $Vol(Y)$ for
its integrated volume.)  Our units are such that the integrated flux is 
\eqn\integratedflux{\mu _{p}\int _YF_{p+2}^{bkgd}\sim \mu _p m^{-(p+1)}Vol(Y)\sim N, }with $\mu _p$ the $p$-brane tension. 
Our particular cases of interest will be IIB on $AdS_5\times Y_5$ and 11d SUGRA
on $AdS_4\times Y_7$, but we'll be more general in this section.  

Metric fluctuations along directions of Killing vectors $K_I^a$ of $Y$ lead to Kaluza-Klein gauge fields $A_I^\mu$ in $M$. Fluctuations of the Ramond-Ramond gauge field background, reduced on non-trivial cycles of $Y$ lead to additional,
``baryonic" gauge fields that we'll also discuss.   In general, Kaluza-Klein reduction involves a detailed, and highly non-trivial, ansatz for how the Kaluza-Klein gauge fields affect the metric and background field strengths.  But here we're simply interested in the 
coefficients $g^{-2}_{IJ}$ of the gauge field kinetic term, and for these it's unnecessary to
employ the full Kaluza-Klein ansatz: a linearized analysis suffices.  

The linearized analysis will be presented in the following section.  In this section, we'll
note some general aspects, and discuss a useful relation that can be obtained by a 
generalization of an argument in \DuffCC, that was based on the non-trivial Kaluza-Klein
ansatz for how the Kaluza-Klein gauge fields modify the backgrounds.  

For Kaluza-Klein isometry gauge fields, both the Einstein term and the $C$ field kinetic terms in
\actfull\ contribute to their gauge kinetic terms:
\eqn\taukkcc{g^{-2}_{IJ}=(g^{-2}_{IJ})^{KK}+(g_{IJ}^{-2})^{CC},}
where $(g^{-2}_{IJ})^{KK}$ is the Kaluza-Klein contribution coming from the Einstein term in 
\actfull\ and $(g^{-2}_{IJ})^{CC}$ is that coming from the Ramond-Ramond $C$ field kinetic terms in \actfull.   On the other hand, if either $I$ or $J$ is a baryonic gauge field, coming from $C$ reduced on a non-trivial cycle of $Y$, then only the $dC$ kinetic terms in \actfull\ contribute  
\eqn\taubbcc{g^{-2}_{IJ}=(g_{IJ}^{-2})^{CC},\qquad\hbox{if $I$ or $J$ is baryonic}.}

Let's review how the Kaluza-Klein contribution in \taukkcc\ is obtained, see e.g. \WeinbergKK. 
Let $y^a$ be coordinates on $Y$, and $K^a_I(y)$ isometric Killing vectors ($I$ labels
the isometry).  The one-form
$d\phi _I$ dual to $K_I$ is shifted by the 1-form gauge field $A_I(x)=A_I^\mu dx^\mu$, with $x^\mu$ coordinates on $M$.  This variation of the metric leads to variation of the Ricci scalar
\eqn\riccivar{R\rightarrow R-{m^{-2}\over 4}g_{ab}(y)K^a_I(y)K^b_J(y)(F_I)_{\mu \nu}(F_J)^{\mu \nu},}
where $ds^2_Y=g_{ab}dy^ady^b$ is the metric on $Y$, with the length
scale $m^{-1}$ factored out.   Since \riccivar\ 
is already quadratic in $A_I$, we don't need to vary $\sqrt{|g|}$.  The contribution to the Kaluza-Klein gauge field kinetic terms coming from the Einstein
action is thus
\eqn\taukk{(g^{-2}_{IJ})^{KK}={m^{-(D_c+2)}\over 16\pi G_{D_t}}\int _{Y}g_{ab}K_I^aK_J^bvol(Y).}
In \WeinbergKK, the Killing vectors are normalized so that the gauge fields have canonical kinetic terms, and then what we're referring to
as the ``coupling" becomes the ``charge" unit; here we'll normalize $K_I^a$ and gauge fields so that the charge
unit is unity, and then physical charges governing interactions are given by what we're calling the couplings $g^{-2}_{IJ}$. 

As an example, it was shown \WeinbergKK\ that reducing the Einstein action on a
$D_c$ dimensional sphere, $Y=S^{D_c}$ of radius $m^{-1}$ leads to 
$SO(D_c+1)$ Kaluza-Klein gauge fields in the uncompactified directions, with coupling 
\WeinbergKK\ 
\eqn\spherew{(g^{-2})^{KK}={1\over 8\pi G_D(D_c+1)m^2}\qquad\hbox{for}\quad Y=S^{D_c},}
with $G_D=G_{D_t}m^{D_c}/Vol(Y)$ the effective Newton's constant in the uncompactified $M_D$.

In   \DuffCC, it was pointed out that \spherew, applied to 11d SUGRA on $S^7$, with 
Freund-Rubin flux for the Ramond-Ramond gauge field, would be incompatible
with the 4d ${\cal N}=8$ $SO(8)$ SUGRA of \deWitEQ, but that properly including the
additional contribution from the Ramond-Ramond fields fixes this problem.  In our
notation above, it was shown in \DuffCC\ that the full coupling of 
 the $SO(8)$
gauge fields in the $AdS_4$ bulk is \eqn\duffr{g^{-2}=(g^{-2})_{KK}+(g^{-2})_{CC}=4g^{-2}_{KK}={1\over 16\pi G_4 m^2},}
which is now perfectly compatible with the 4d ${\cal N}=8$ theory of \deWitEQ. 

We here point out that, for general Freund-Rubin
compactifications on any Einstein space $Y$ of dimension $D_c$, there is always a fixed proportionality
between the Einstein and Ramond-Ramond contributions to the Kaluza-Klein
gauge kinetic terms: 
\eqn\taukkcc{(g^{-2}_{IJ})^{CC}={D_c-1\over 2}(g^{-2}_{IJ})^{KK},}
of which \duffr\ is a special case.  Our relation \taukkcc\ follows from a generalization of the argument in \DuffCC.  In a KK ansatz like that of \duffr, the contribution to $g^{-2}_{IJ}$ from
the Ramond-Ramond kinetic term in \actfull\ is 
\eqn\rrktis{(g_{IJ}^{-2})^{CC}={m^{-(D_c+2)}\over 16\pi G_{D_t}} \int _Y{1\over 2}g_{ab}\grad c K^a_I\grad {}^c K^b_Jvol(Y)={D_c-1\over 2}(g^{-2}_{IJ})^{KK}.}
In the last step, there was an integration by parts, use of  $-\grad c\grad {}^cK_I^a= R_c^aK_I^c$, use of $R_{ab}=(D_c-1)m^2g_{ab}$ since $Y$ is
taken to be Einstein, and comparison with 
\taukk.   We will check and verify the relation \taukkcc\ more explicitly in the following sections.

As a quick application, we find from \spherew\ and \taukkcc\ that reducing 10d IIB SUGRA
on $S^5$ leads to a theory in the $AdS_5$ bulk with $SO(6)$ gauge fields with coupling 
\eqn\sosix{g^{-2}_{SO(6)}=(g^{-2}_{SO(6)})^{KK}+(g^{-2}_{SO(6)})^{CC}=3(g^{-2}_{SO(6)})^{KK}={L^2\over 16\pi G_5},}
where $m^{-1}=L$ is the radius of the $S^5$, and also the length scale of the $AdS_5$ vacuum.  The result 
\sosix\  agrees with that found in \GRW\ for 5d $\N =8$ SUGRA: the $SO(5)$ 
invariant vacuum in eqn. (5.43) of \GRW\ has, in $4\pi G_5=1$ units,  $R_{\mu \nu}=g^2g_{\mu \nu}$;  thus  $g^{-2}=L^2/4=L^2/16\pi G_5$, in agreement with \sosix.
 Using \kappavis, with $Vol(S^5)=\pi ^3$, gives $\tau _{SO(6)}=8\pi ^2Lg^{-2}=\pi L^3/2G_5=N^2$.  On the other hand, \taurr\ here gives $\tau _{RR}=4N^2/3$.  We
can also verify $\tau _{RR}=4N^2/3$ by direct computation in the $\N =4$ theory (where
the free field value is not renormalized).  The apparent difference with the above
$\tau _{SO(6)}$ is because of the different normalization of the $U(1)_R$ vs. $SO(6)$
generators. 

The relation \taukkcc\ will prove useful in what follows, because the Ramond-Ramond
contribution $(g^{-2}_{IJ})^{CC}$ is sometimes, superficially, easier to compute than 
the Kaluza-Klein contribution \taukk.  Thanks to the general
relation \taukkcc, the full coefficient of the kinetic terms for Kaluza-Klein gauge fields can be
computed from $(g^{-2}_{IJ})^{CC}$ as
\eqn\tauisi{g^{-2}_{IJ}=(g^{-2}_{IJ})^{KK}+(g^{-2}_{IJ})^{CC}=
{D_c+1\over D_c-1}(g^{-2}_{IJ})^{CC}.}

\newsec{Gauge fields and associated $p$-forms on $Y$}

The linearized fluctuations of the gauge fields modify the background as
\eqn\fbacks{F_{p+2}^{bkgd}\rightarrow (p+1)m^{-(p+1)}vol(Y)+d \left(\sum _I  \omega _I
\wedge A_I \right) ,}
and hence, writing $F=dC$, 
\eqn\cbacks{C_{p+1}\rightarrow C_{p+1}^{bkgd}+\sum _I \omega _I\wedge A_I}
Here $A_I$ are all of the gauge fields, both Kaluza-Klein and the baryonic ones coming
{}from reducing $C_{p+1}$ on non-trivial $p$ cycles of $Y$.  

So every gauge field $A_I$ enters into $C_{p+1}$ at the linearized level, and we'll here
be interested in determining the associated form $\omega _I$ in \cbacks.  
 The $\omega _I$ associated with Kaluza-Klein gauge fields $A_I$ 
are found from the variation of $vol(Y)$ in \fluxgen\ by the linearized shift of the 1-form, 
dual to the Killing vector isometry $K_I$, by $A_I$:
\eqn\volvar{vol(Y)\rightarrow vol(Y)+d\left(\sum _I \widehat \omega _I\wedge A_I\right) , \qquad\hbox{with}\qquad d\widehat \omega _I=i_{K_I}vol(Y).}
Using this in \fbacks\ gives \cbacks, with 
associated $p$-form $\omega _I \equiv (p+1)m^{-(p+1)}\widehat \omega _I$ on $Y$.

Note that this definition of the $\omega _I$ is ambiguous under shifts of the $\omega _I$
by any closed $p$ form.  Shifts of $\omega _I$ by any exact form will have no effect, so
this ambiguity in defining the $\omega _I$ associated with Kaluza-Klein gauge fields
is associated with the cohomology $H_p(Y)$ of closed, mod exact, $p$ forms on $Y$.

The baryonic gauge fields $A_I$ enter into \cbacks\ with $\omega _I$
running over a basis of the cohomology $H_p(Y)$ of closed, mod exact, $p$-forms on $Y$.
The ambiguity mentioned above in the Kaluza-Klein gauge fields corresponds to the
freedom in one's choice of basis of the global symmetries, as any linear combination of
a ``mesonic" flavor symmetry and any ``baryonic" flavor symmetry is also a valid ``mesonic" 
flavor symmetry.  

Branes that are  electrically charged under $C_{p+1}$ have worldvolume coupling $\mu _p \int   C_{p+1}$, with $\mu _p$ the brane tension.  Wrapping these branes on
the non-trivial cycles $\Sigma$ of $H^p(Y)$ yield particles in the uncompactified dimensions,
and \cbacks\ implies that these wrapped branes carry electric charge
\eqn\wrpcrg{q_I(\Sigma)=\mu _p\int _\Sigma \omega _I}
under the gauge field $A_I$.  

Plugging \cbacks\ into $F_{p+2}$ kinetic terms in \actfull\ gives
what we called the $(g^{-2}_{IJ})^{CC}$ contribution to the gauge field kinetic terms to
be 
\eqn\gccis{(g^{-2}_{IJ})^{CC}={1\over 16\pi G_{D_t}}\int _Y \omega _I\wedge *\omega _J
\equiv {(p+1)^2m^{-(p+4)}\over 16\pi G_{D_t}}\int _Y\widehat \omega _I\wedge *\widehat \omega _J,}
where $\omega _I\equiv (p+1)m^{-(p+1)}\widehat \omega _I$ and $*\omega _I \equiv 
(p+1)m^{-3}*\widehat \omega _I$ 

We will use \gccis, together with \tauisi\ for Kaluza-Klein gauge fields, or \taubbcc\ for baryonic gauge fields, to compute the coefficients $g^{-2}_{IJ}$ of the gauge field kinetic terms
in $AdS_{d+1}$.  These are then related to the coefficients, $\tau _{IJ}$, of the current-current
two-point functions in the gauge theory according to \tauads.

\newsec{Sasaki-Einstein $Y$, and the form $\omega _R$ for the R-symmetry.}

The modification \cbacks\ for the $U(1)_R$ gauge field, coming from the $U(1)_R$ isometry
of Sasaki-Einstein spaces, was found in \BHK, which we'll review in this section.  

The metric of Sasaki-Einstein $Y_{2n-1}$ can locally be written as
\eqn\seyv{ds^2(Y)=({1\over n}d\psi '+\sigma)^2+ds_{2(n-1)}^2,}
with $ds_{2(n-1)}^2$ a local, Kahler-Einstein metric, and
\eqn\jandom{d\sigma = 2J\qquad d\Omega =ni\sigma \wedge \Omega ,}
with $J$ the local Kahler form and $\Omega $ the local holomorphic $(n-1,0)$ form
for $ds_{2(n-1)}^2$.  In \BHK\ the coordinate $\psi =\psi '/q$ was used, in order to have the range
$0\leq \psi <2\pi$; $q$ is given by $nd\sigma = 2\pi q c_1$, with $c_1$ the first
Chern class of the $U(1)$ bundle over the $n-1$ complex dimensional Kahler-Einstein
space with metric $ds_{2(n-1)}^2$.  The $U(1)_R$ isometry is associated with the Reeb
Killing vector
\eqn\reebv{K=n{\partial \over \partial \psi '}.}
It is convenient to define the unit 1-form, dual to the Reeb vector, of the $U(1)_R$ fiber
\eqn\epsi{e^\psi \equiv {1\over n}d\psi ' + \sigma.}
Note that $de^\psi = d\sigma =2J$. 
The volume form of $Y_{2n-1}$ is
\eqn\volyiv{vol(Y_{2n-1})={1\over (n-1)!} e^\psi \wedge J^{n-1}.}

Following \BHK, the linearized effect of the $U(1)_R$ isometry \reebv\ Kaluza-Klein gauge field is found by shifting
\eqn\epsishift{e^\psi \rightarrow e^\psi +{2\over n}A_R,}
where the coefficient of $A_R$ is chosen so that the $U(1)_R$ symmetry is properly
normalized: the holomorphic n-form on $C(Y)$, which leads to superpotential terms, has
R-charge 2.  
The shift \epsishift\ affects the volume form \volyiv\ as
\eqn\volsh{vol(Y_{2n-1})\rightarrow vol(Y_{2n-1})+{2\over n!}A_R\wedge J^{n-1}-{1\over n!}dA_R\wedge e^\psi \wedge J^{n-2},}
where the last term in \volsh\ was added to keep the form closed:
\eqn\volshx{vol(Y_{2n-1})\rightarrow vol(Y_{2n-1})+d\left({1\over n!}e^\psi \wedge J^{n-2}\wedge A_R\right).}

The shift \volshx\ alters the Ramond-Ramond flux background $F^{bkgd}_{2n-1}$ \fbacks,
and thus alters $C_{2n-2}$ as in \cbacks, $\delta C_{2n-2}= \omega _R\wedge A_R$, with the $2n-3$ form  $\omega _R$ given by
\eqn\omegario{\widehat \omega _R \equiv {\omega _R \over (2n-2)m^{-(2n-2)}}={1\over n!}
e^\psi \wedge J^{n-2}.}

In particular, for type IIB on $AdS_5\times Y_5$, the background flux is 
\eqn\civbkgdx{F_5^{bkgd}=4L^4\left(vol(Y_5)+ *vol(Y_5)\right),}
and \volshx\ alters the $C_4$ on $Y_5$ as in \cbacks, 
 with 3-form $\omega _R$ given by  \BHK:
\eqn\omegaris{\widehat \omega _R\equiv {1\over 4L^4}\omega _R ={1\over 6}e^\psi \wedge J, \qquad \hbox{for $Y_5$}.}

For 11d SUGRA on $AdS_4\times Y_7$, the effect
of \volshx\ on the Ramond-Ramond flux 
\eqn\cvibkgd{F_7=6(2L)^6vol(Y_7)}
leads to a shift as in \cbacks\ of $C_6$, by $\omega _R\wedge A_R$, with 5-form 
 $\omega _R$ given by \BHK
\eqn\omvr{\widehat \omega _R\equiv {1\over 6(2L)^6}\omega _R={1\over 24}e^\psi \wedge J\wedge J.}

Wrapping a brane on a supersymmetric $2n-3$ cycle $\Sigma$ of $Y$ yields a baryonic particle $B_\Sigma$ in the $AdS_{d+1}$ bulk, dual to a baryonic chiral operator in the gauge theory. It was verified in \BHK\ that the R-charges assigned to such objects by the forms
\omegaris\ and \omvr\ are compatible with the relation \delR\ in the dual field theory.
Using \omegario, the R-charge assigned to such an object is related to the operator dimension $\Delta$ as 
\eqn\rsigma{\eqalign{R[B_\Sigma]&=\mu _{2n-3}\int _{\Sigma _{2n-3}}\omega _R= {2\over n}\mu _{2n-3}m^{-(2n-2)}\int _\Sigma {1\over (n-2)!}e^\psi \wedge J^{n-2}\cr &={2\over n}\mu _{2n-3}m^{-(2n-2)}Vol(\Sigma _{2n-3})={2m^{-1}\over nL}  \Delta [B_\Sigma].}}
In going from the first to the second line of \rsigma, we used the fact that the supersymmetric
$2n-3$ cycles in $Y$ are calibrated, with $vol(\Sigma )=e^\psi \wedge J^{n-2}/(n-2)!$. 
For both IIB on $AdS_5\times Y_5$ and M theory on $AdS_4\times Y_7$, \rsigma\ matches
with the relation \delR\ in the 4d and 3d dual, respectively \BHK: in the former case, 
$m^{-1}=L$ and $n=3$ in \rsigma, and in the latter case $m^{-1}=2L$ and $n=4$.

The $\mu _{2n-3}m^{-(2n-2)}$ factor in \rsigma\ is proportional to $N/Vol(Y)$ by
the flux quantization condition.  For $AdS_5\times Y_5$, using \fluxq\ then gives \BHK
\eqn\rvolv{R(\Sigma _i)={2\over 3}\mu _3 L^4 Vol(\Sigma _i)={\pi N\over 3}{Vol(\Sigma _i)\over Vol(Y _5)}.}
For  M theory on $AdS_4\times Y_7$, the flux quantization condition (see e.g. the recent work \GLMW)
\eqn\fqvii{6(2L)^6Vol(Y_7)=(2\pi \ell _{11})^6N,}
where $16\pi G_{11}=(2\pi )^8\ell _{11}^9$.  Using the M5 tension $\mu _5=1/(2\pi )^4\ell _{11}^6$,   \rsigma\ then gives
\eqn\rvolvii{R(\Sigma _i)={\pi ^2 N\over 3}{Vol(\Sigma _i)\over Vol(Y _7)}.}

\newsec{The forms $\omega _I$ for other symmetries}

In this section, we find the forms entering in \cbacks, for the non-R flavor symmetries.  Those
associated with non-R isometries are found in  direct analogy with the
discussion of \BHK, reviewed in the previous section, for $\omega _R$.  We re-write
\volyiv\ as
\eqn\volyivx{vol(Y_{2n-1})={1\over 2^{n-1}(n-1)!} e^\psi \wedge (de^\psi)^{n-1}.}
Under a non-R isometry, the form $e^\psi$ \epsi\ shifts by 
\eqn\epsifs{e^\psi \rightarrow e^\psi +h_i(Y)A_{F_i},}
with the  functions $h_i(Y)$ obtained by contracting the 1-form $\sigma$ in
\epsi\ with the Killing vector $K_i$ for the flavor symmetry,
\eqn\hiis{h_i(Y)=i_{K_i}\sigma = g_{ab}K^aK_i^b.}
The last equality follows from \seyv: $i_{K_i}\sigma$ can be obtained by contracting the
Reeb vector $K^a$ and the general Killing vector $K_i^b$, using the metric \seyv. 

In the last section, for $U(1)_R$, only the first $e^\psi$ factor in \volyivx\ was shifted,
as that $e^\psi$ factor is associated with the $U(1)_R$ fiber, where $U(1)_R$ acts.
Conversely, since non-R isometries do not act on the $U(1)_R$ fiber, but rather
in the Kahler Einstein base, we should not shift the first $e^\psi$ factor in \volyivx,
but instead shift the $n-1$ factors of $de^\psi$ in \volyivx.   Effecting this shift
gives
\eqn\volyivxv{\delta vol(Y_{2n-1})={1\over 2^{n-1}(n-2)!}\left(e^\psi \wedge 
 d(h_i(Y)A_{F_i})\wedge (de^\psi)^{n-2} -de^\psi \wedge 
h_i(Y)A_{F_i}\wedge (de^\psi)^{n-2}\right),}
where the last term was added to keep the form closed:
\eqn\volyivxvx{\delta vol(Y_{2n-1})=-d\left({1\over 2(n-2)!}h_i(Y)e^\psi \wedge 
J^{n-2}\wedge A_{F_i}\right).}
Effecting this shift in $F^{bkgd}$ leads 
 to $\delta C_{2n-2}=\omega _{F_i}\wedge A_{F_i}$, with $2n-3$ form $\omega _{F_i}$:
\eqn\omegafio{\widehat \omega _{F_i}\equiv {\omega _{F_i}\over (2n-2)m^{-(2n-2)}}=-{1\over 2(n-2)!}h_i(Y)e^\psi \wedge J^{n-2}=-{n(n-1)\over 2}h_i(Y)\widehat \omega _R.}
 Aside from the factor of $-\half n(n-1)h_i(Y)$,
$\omega _{F_i}$ is the same as for $\omega _R$, as given in \omegario.

In particular, for IIB on $AdS_5\times Y_5$ we have 
\eqn\omegafis{\widehat \omega _{F_i}\equiv {\omega _{F_i}\over 4L^4}
=-{1\over 2}h_i(Y_5)e^\psi \wedge J=-3 h_i(Y_5)\widehat \omega _R,}
and for M theory on $AdS_4\times Y_7$ we have 
\eqn\omegafiss{\widehat \omega _{F_i}\equiv {1\over 6(2L)^6} \omega _{F_i}=-{1\over 4}h_i(Y_7)e^\psi \wedge  J\wedge J= -6 h_i(Y_7)\widehat \omega _R.}

As reviewed in \rsigma, the R-charge of branes wrapped on supersymmetric cycles 
$\Sigma$ is
\eqn\rsigmagain{R[B_\Sigma ]={2\over n}\mu _{2n-3}m^{-(2n-2)}\int _\Sigma vol(\Sigma).}
Using \omegafio, the flavor charges of these wrapped branes can similarly be written as
\eqn\fsigma{\eqalign{F_i[B_\Sigma]&=\mu _{2n-3}\int _\Sigma \omega _{F_i}=-(n-1)\mu _{2n-3}m^{-(2n-2)}\int _\Sigma h_ivol(\Sigma)\cr &=
-{n(n-1)\over 2}\cdot R[B_{\Sigma }]\cdot {\int _{\Sigma}h_i vol(\Sigma)\over \int _\Sigma vol(\Sigma )}.} }
In particular, for IIB on $AdS_5\times Y_5$, we have
\eqn\fsigmagain{F_i[B_\Sigma ]=
-{\pi N \over Vol(Y)}\int _{\Sigma _3}h_ivol(\Sigma)=-3R[B_\Sigma]{\int _\Sigma h_i vol(\Sigma)\over \int _\Sigma vol(\Sigma)}.}

The baryonic symmetries, coming from reducing $C_{2n-2}$ on the
non-trivial $(2n-3)$-cycles of $Y_{2n-1}$, also alter $C_{2n-2}$ at linear order as in \cbacks,
$\delta C_{2n-2}=\omega _{B_i}\wedge A_{B_i}$, 
where the $2n-3$ forms $\omega _{B_i}$ are representatives of the cohomology
$H_{2n-3}(Y,\IZ )$.   These can be locally written on $Y_{2n-1}$ as 
\eqn\omegabis{\omega _{B_i}=k_i e^\psi \wedge \eta _i,}
where $\eta _i$ are $2(n-2)$ forms on the Kahler-Einstein base, satisfying
$d\eta _i=0$, and $\eta _i\wedge J=0$.  The normalization constants $k_i$ in  \omegabis\ are chosen so that $\mu _{2n-3}\int _{\Sigma }\omega _{B_i}$ is an integer for all $(2n-3)$-cycles
$\Sigma$ of $Y_{2n-1}$.  

As mentioned in sect. 4, this construction of the forms 
$\omega _{F_i}$ involves integrating an 
expression for $d\omega _{F_i}$, so there's an ambiguity of 
adding an arbitrary closed form
to $\omega _{F_i}$.  Since addition of an exact form would not affect the charges of branes
wrapped on closed cycles, the interesting ambiguity corresponds precisely to the same cohomology class of forms as the $\omega _{B_j}$.  This is as it should be: 
there is an ambiguity in our basis for the mesonic flavor symmetries, as one can always re-define them by arbitrary additions of the baryonic flavor 
symmetries.   The form \omegafio\ for $\omega _{F_i}$ corresponds to some particular 
choice of the basis for the mesonic flavor symmetries.  In the field theory dual, it may look 
more natural to call this a linear combination of mesonic and baryonic flavor symmetries.  

\newsec{Computing $\tau _{IJ}$ from the geometry of $Y$}

The expressions \gccis\ 
for the Ramond-Ramond kinetic term
contribution $(g^{-2}_{IJ})^{CC}$ is 
\eqn\gccisx{(g^{-2}_{IJ})^{CC}={1\over 16\pi G_{D_t}}\int _Y \omega _I\wedge *\omega _J
\equiv {(2n-2)^2m^{-(2n+1)}\over 16\pi G_{D_t}}\int _Y\widehat \omega _I\wedge *\widehat \omega _J}
and the Einstein action contribution \taukk\ is 
\eqn\taukkx{(g^{-2}_{IJ})^{KK}={m^{-(2n+1)}\over 16\pi G_{D_t}}  \int _{Y_{2n-1}}g_{ab}K_I^aK_J^b vol(Y_{2n-1});}
again, the length scale $m^{-1}$ is factored out of the metric and volume form.
As discussed in sect. 3, for gauge fields associated with isometries of $Y$, and in particular
the graviphoton, we add the two contributions, $g^{-2}_{IJ}=(g^{-2}_{IJ})^{CC}+(g^{-2}_{IJ})^{KK}$, whereas for baryonic symmetries there is no contribution from the Einstein
action, so $g^{-2}_{IJ}=(g^{-2}_{IJ})^{CC}$.

Our claimed general proportionality \taukkcc\ here gives
\eqn\rrktiso{(g^{-2}_{IJ})^{CC}=(n-1)(g^{-2}_{IJ})^{KK},}
which implies that 
\eqn\rrktisx{4(n-1)\int _{Y_{2n-1}} \widehat \omega _I\wedge *\widehat \omega _J= \int _{Y_{2n-1}}
g_{ab}K^a_IK^b_J vol(Y_{2n-1}).}
As we'll see, this relation can look non-trivial in the geometry.

To compute $(g^{-2}_{IJ})^{CC}$ from \gccisx, we first note that \omegario\ gives
\eqn\omegariod{*\widehat \omega _R\equiv {*\omega _R\over (2n-2)m^{-3}}=
{1\over n!}*e^\psi \wedge J^{n-2}={n-2\over n!} J,}
and then, using \volyiv, gives
\eqn\omromrd{\widehat \omega _R \wedge *\widehat \omega _R={(n-2)\over n!n}vol(Y_{2n-1}).}

In particular, for the $U(1)_R$ graviphoton, we obtain
\eqn\grrccgen{(g^{-2}_{RR})^{CC}= {(2n-2)^2m^{-(2n+1)}\over 16\pi G_{D_t}}{(n-2)\over n!n}Vol(Y_{2n-1}).}

For the mixed kinetic term between $U(1)_R$ and non-R isometries $U(1)_{F_i}$, 
\eqn\grfccgen{(g^{-2}_{RF_i})^{CC}= {(2n-2)^2m^{-(2n+1)}\over 16\pi G_{D_t}}{(n-2)\over n!n}\left(-{n(n-1)\over 2}\right)\int _Y h_i(Y)vol(Y).}
For the $U(1)_{F_i}$ and $U(1)_{F_j}$ kinetic terms, we similarly obtain
\eqn\gffccgen{(g^{-2}_{F_iF_j})^{CC}= {(2n-2)^2m^{-(2n+1)}\over 16\pi G_{D_t}}{(n-2)\over n!n}\left({n(n-1)\over 2}\right)^2\int _Y h_i(Y)h_i(Y)vol(Y).}

For $U(1)_{B_i}$ symmetries, we have
\eqn\gbrgen{g^{-2}_{RB_i}={1\over 16\pi G_{D_t}}\int _Y\omega _{B_i}\wedge *\omega _R={(2n-2)m^{-(2n-2)}\over 16\pi G_{D_t}}{n-2\over n!}\int _Yk_ie^\psi \wedge\eta _i \wedge J=0,}
where we used \omegabis\ for $\omega _{B_i}$, \omegariod, and we get zero immediately from $\eta _i \wedge J=0$.  Likewise, 
\eqn\gbfgen{g^{-2}_{F_jB_i}=0,}
for any isometry symmetry $F_i$, since \omegafio\ gives $\omega _{F_j}\propto \omega _R$,
so $*\omega _{F_i}\propto J$, and we immediately get zero in \gbfgen\ again from 
$\eta _i\wedge J=0$.  As mentioned in the introduction, there is thus never any
kinetic term mixing between any of the isometry Kaluza-Klein gauge fields and any of the
gauge fields coming from reducing the $C$ fields on non-trivial homology cycles of $Y$. 
Finally, for the baryonic kinetic terms, we have
\eqn\gbrgen{g^{-2}_{B_iB_j}={1\over 16\pi G_{D_t}}\int _Yk_ik_j e^\psi \wedge \eta _i \wedge *_B\eta _j,}
where $*_B$ acts on the $2n-2$ dimensional Kahler-Einstein base.  

For the isometry (non-baryonic) gauge fields, we have to add the Kaluza-Klein contributions,
$(g^{-2}_{IJ})^{KK}$, from the
Einstein action, to the kinetic terms.  These can either be explicitly computed, using \taukkx,
or one can just use our relation \rrktisx\ to the above Ramond-Ramond contributions.
It's interesting to check that our relation \rrktisx\ is indeed satisfied.  
For example, the Kaluza-Klein contribution $(g^{-2}_{RR})^{KK}$ is 
\eqn\grrkkgen{{m^{-(2n+1)}\over 16\pi G_{D_t}}  \int _{Y_{2n-1}}g_{ab}K^aK^b vol(Y_{2n-1})=  {m^{-(2n+1)}\over 16\pi G_{D_t}} {4\over n^2}Vol(Y_{2n-1}),}
where we used the local form of the metric \seyv, and $U(1)_R$ isometry Killing vector \reebv, rescaled
by the factor in \epsishift\ to have $U(1)_R$ properly normalized.  Comparing with 
\grrccgen, our relation \rrktisx\ is indeed satisfied for both of our cases of interest, 
$n=3$ and $n=4$, appropriate for IIB on $AdS_5\times Y_5$ and M theory on
$AdS_4\times Y_7$, respectively.  

Our main point will be that the $\tau _{R_tR_t}$ minimization condition \taufrs\ of 
\taumin\ requires \grfccgen\ to vanish, $\tau _{RF_i=0}$, so we must have
\eqn\hintz{\int _Y h_i(Y)vol(Y)=\int _Yi_{K_i}\sigma vol(Y)=\int _Y g_{ab}K^aK_i^b=0,}
for every non-R isometry Killing vector $K_i^a$.
 We know from the field theory argument of
\taufrs\ that the conditions \hintz\ must uniquely determine which, among all possible
R-symmetries, is the superconformal R-symmetry.  Correspondingly, \hintz\ determines
the isometry $K$, from among all possible mixing with the $K_i^a$.  As we'll discuss in the following
sections, the Z-minimization of \MSY\ precisely implements \hintz\ (in the context
of toric $C(Y)$).  Also, \gbrgen\ 
implies that the condition
$\tau _{Ri}$ of \taumin\ is automatically satisfied for baryonic $U(1)_{B_i}$.
This is the reason why the Z-minimization method of \MSY\ did not need to include any mixing
of $U(1)_R$ with the baryonic $U(1)_B$ symmetries.

For future reference, we'll now explicitly write out the above formulae for our cases of
interest.  For IIB on $AdS_5\times Y_5$, we have $n=3$ and $m^{-1}=L$, so \gccisx\ is
\eqn\tauadsv{\tau _{IJ}^{CC}\equiv 8\pi ^2 L(g^{-2}_{IJ})^{CC}={8\pi L^8\over G_{10}}\int _{Y_5}\widehat \omega _I\wedge *\widehat \omega _J={16N^2\pi ^3\over Vol(Y_5)^2} 
\int  _{Y_5}\widehat \omega _I\wedge * \widehat \omega _J,}
where we used \fluxq\ to write the result in terms of $N$.   For $I$ or $J$ baryonic, this
is the entire contribution:
\eqn\taubar{\tau _{IJ}={16N^2\pi ^3\over Vol(Y_5)^2}\times 
\int  _{Y_5}\widehat \omega _I\wedge *\widehat \omega _J, \qquad \hbox{for $I$ or $J$
baryonic}.}
For isometry gauge fields, we add this to 
\eqn\tauadsvkk{\tau _{IJ}^{KK}={8\pi ^2L^8\over 16\pi G_{10}}\int _{Y_5}vol(Y_5) g_{ab}K_I^aK_J^b={N^2\pi ^3\over Vol(Y_5)^2}\int _{Y_5}vol(Y_5) g_{ab}K_I^aK_J^b,}
or, using relation \taukkcc, we simply have 
\eqn\tauisixx{\tau _{IJ}={3\over 2}\tau _{IJ}^{CC}={24N^2\pi ^3\over Vol(Y_5)^2}
\int  _{Y_5}\widehat \omega _I\wedge *\widehat \omega _J,\qquad\hbox{for $I$ and $J$ Kaluza-Klein.}}

In particular, for the $U(1)_R$ kinetic term we compute  
\eqn\taurrcc{\tau _{RR}^{CC}={16N^2\pi ^3\over Vol(Y_5)^2}\int _{Y_5}\omega _R\wedge *\omega _R={16N^2\pi ^3\over Vol(Y_5)^2}\int _{Y_5}{1\over 36}e^\psi \wedge J\wedge J=
{8N^2\pi ^3\over 9 Vol(Y_5)},}
and 
\eqn\taurrkkis{\tau _{RR}^{KK}={N^2\pi ^3\over Vol(Y_5)^2}\int _{Y_5}{4\over 9}vol(Y_5)= {4N^2\pi ^3\over 9Vol(Y_5)},}
verifying \taukkcc. 
The total for the graviphoton kinetic term coefficient then gives
\eqn\taurrtot{\tau _{RR}=\tau _{RR}^{CC}+\tau _{RR}^{KK}={4\over 3}{N^2\pi ^3\over Vol(Y_5)}.}
This agrees perfectly with the relation \taurrct\ and \susyreln, given \hsac. 

For the kinetic terms for two mesonic non-R symmetries, \tauisixx\ gives 
\eqn\tauffx{\tau _{F_iF_j}={12N^2\pi ^3\over Vol(Y_5)^2}\times \int _{Y_5}h_ih_jvol(Y_5).}
The relation \taukkcc, $\tau _{IJ}^{KK}=\half \tau _{IJ}^{CC}$, which was already used in \tauffx\ can be written as 
\eqn\taukkccff{ \int _{Y_5}
g_{ab}K_{F_i}^aK^b_{F_j}vol(Y_5)=4 \int _{Y_5}h_ih_jvol(Y_5)=4\int _{Y_5}g_{ac}g_{bd}K^cK^d K_{F_i}^aK_{F_j}^bvol(Y_5).}

Likewise, using \taubar, the kinetic terms for two baryonic flavor symmetries are
\eqn\taubbx{\tau _{B_iB_j}={16N^2\pi ^3\over Vol(Y_5)^2}k_i  k_j \int _{Y_5}e^\psi \wedge \eta _i \wedge *(e^\psi \wedge \eta _j).}

For $M$ theory on  $AdS_4\times Y_7$, we set $n=4$ for $Y_7$, and $m^{-1}=2L$ for its length scale, in the above expressions.  Then we obtain from \gccisx, using also \tauads\ with $d=3$, 
\eqn\tauadsvii{\tau _{IJ}^{CC}\equiv 4\pi  (g^{-2}_{IJ})^{CC}=4\pi (6)^2(2L)^{9}{1\over 16\pi G_{11}}\int \widehat \omega _I\wedge *\widehat \omega _J.}
Using the flux quantization relation \fqvii, \tauadsvii\ becomes
\eqn\tauadsviix{\tau _{IJ}^{CC}={48\pi ^2N^{3/2}\over \sqrt{6}(Vol(Y_7))^{3/2}}\int _{Y_7}\widehat \omega _I\wedge \widehat \omega_J.}
 Using \taukk\ we can also write the Kaluza-Klein contribution, as
\eqn\tauadsviikk{\tau ^{KK}_{IJ}\equiv 4\pi (g^{-2}_{IJ})^{KK}= {4\pi ^2 N^{3/2}\over 3\sqrt{6}
(Vol(Y_7))^{3/2}} \int _{Y_7}g_{ab}K_I^aK_J^b vol(Y_7) .}

For $\tau _{RR}$, \grrccgen\ gives 
\eqn\taurrccvii{\tau _{RR}^{CC}={\pi ^2N^{3/2}\over \sqrt{6Vol(Y_7)}}.}
The Kaluza-Klein contribution is given by \taukkx, with 
$g_{ab}K_R^aK_R^b=(1/2)^2$ from \epsishift, so
\eqn\taurrkkvii{\tau _{RR}^{KK}={\pi ^2 N^{3/2}\over 3\sqrt{6
Vol(Y_7)}} .}
Comparing \taurrccvii\ and \taurrkkvii, we verify that $\tau _{RR}^{CC}=3\tau _{RR}^{KK}$,
in agreement with our general expression \rrktis\ (specializing $Y_7=S^7$ gives the case analyzed in \DuffCC).  The total is
\eqn\taurrvii{\tau _{RR}=  {4\pi ^2 N^{3/2}\over 3\sqrt{6
Vol(Y_7)}} .}

We can compare \taurrvii\ with the 3d $\N =2$ gauge theory proportionality relation
\eqn\taurrciii{\tau _{RR}={\pi ^3\over 3}C_T\qquad\hbox{in $d=3$,}}
where $C_T$ is the coefficient of the stress tensor two-point function.  
Along the lines of \refs{\MHKS, \GubserVD},  the central charge $C_T$ is determined 
in the dual, from
the Einstein term of $M$ theory on $AdS_4\times Y_7$, to be 
\eqn\ctiiig{C_T={(2N)^{3/2}\over \pi \sqrt{3Vol(Y_7)}},}
so \taurrvii\ indeed satisfies \taurrciii.
 As a special case,
for $Y_7=S^7$, $Vol(S^7)=\pi ^4/3$ and \taurrvii\ gives $\tau _{RR}=(2N)^{3/2}/3$. 

For two non-R isometries , we have from \tauadsvii\ and \rrktiso, for 
 $AdS_4\times Y_7$:
\eqn\taufifjvii{\tau _{F_iF_j}={4\over 3}\tau _{F_iF_j}^{CC}={\pi ^2 (2N)^{3/2}\over 3\sqrt{3}(Vol(Y_7))^{3/2}}\int _{Y_7}(6)^2
h_ih_j vol(Y).}

\newsec{Toric Sasaki-Einstein Geometry and Z-minimization}

In this section, we'll briefly summarize some of the results of \MSY.  Consider a Sasaki-Einstein manifold $Y_{2n-1}$, of real dimension $2n-1$, whose metric cone $X=C(Y)$
\xandh\ 
is a local Calabi-Yau $n$-fold.  The condition that \xandh\ be Kahler is equivalent to $Y=X|_{r=1}$ 
being Sasaki, which is needed for the associated field theory to be supersymmetric.  The complex structure of $X$ pairs the Euler vector $r\partial /\partial r$ with the Reeb vector $K$, $K={\cal I}(r\partial /\partial r)$.  This is the AdS dual version of the pairing, by supersymmetry, between the dilitation 
generator and the superconformal R-symmetry, respectively.  The physical problem of determining
the superconformal R-symmetry among all possibilities \rgen\ maps to the mathematical problem of
determining the Reeb vector among all $U(1)$ isometries of $Y$.  

When  $X=C(Y)$ is toric, it can be given local coordinates $(y^i, \phi _i)$, $i=1\dots n$, and both $C(Y)$ and $Y$ have a $U(1)^n$ isometry group, associated with the torus coordinates $\phi _i\sim \phi _i +2\pi$.  It is useful
to introduce both symplectic coordinates $(y^i, \phi _i)$ and complex coordinates $(x_i, \phi _i)$.  In the symplectic coordinates, the 
symplectic Kahler form is simply $\omega = dy^i\wedge d\phi _i$, and the 
metric with toric $U(1)^n$ isometry takes the form
\eqn\metsymp{ds^2=G_{ij}dy^idy^j+G^{ij}d\phi _id\phi _i,}
with $G^{ij}$ the inverse to $G_{ij}(y)$, and $G_{ij}=\partial ^2G/\partial y^i\partial y^j$ for some convex symplectic potential function $G(y)$.  In the complex
coordinates,  $z_i=x_i+i\phi _i$, the metric is 
\eqn\metcomp{ds^2=F^{ij}dx_idx_j+F^{ij}d\phi _id\phi _i,}
and $F^{ij}=\partial ^2F(x)/\partial x_i\partial x_j$, with $F(x)$ the Kahler potential.  The two coordinates are related by a Legendre 
transform, $y^i=\partial F(x)/\partial x _i$ and $F^{ij}(x)=G^{ij}(y=\partial F/\partial x)$, with  
$F(x)=(y_i \partial G/\partial y_i-G)(y)$.  The holomorphic n-form of the cone $X=C(Y)$ is
\eqn\holonis{\Omega _n=e^{x_1+i\phi _1}(dx_1+id\phi _1)\wedge \dots \wedge (dx_n+id\phi _n).}

The
Reeb vector can be expanded as 
\eqn\Kb{K=b_i{\partial \over \partial \phi _i},}
and its symplectic pairing with $r{\partial \over \partial r}$ implies that 
\eqn\bgy{b_i=2G_{ij}y^j, \qquad \hbox{note:}\quad b_i=\hbox{constant}.}
The problem of determining the superconformal R-symmetry maps to that of determining the
coefficients $b_i$, $i=1\dots n$.  The component $b_1$ is fixed to $b_1=n$ by the condition that ${\cal L}_K\Omega _n=in\Omega _n$, which is the condition that $U(1)_R$ in the field
theory is properly normalized to give the superpotential charge $R(W)=2$. 
The remaining $n-1$ components $b_i$ are unconstrained by symmetry conditions,
corresponding to the field theory statement that $U(1)_R$ can mix with an $U(1)^{n-1}$ group of non-R 
flavor symmetries.  

The space $X=C(Y)$ is mapped by the moment map, $\mu$, where one forgets the
angular coordinates $\phi _i$, to ${\cal C}=\{ y | (y, v_a)\geq 0\}$, where $v_a\in \IZ ^n$, for
$a=1\dots d$, are the ``toric data".  The supersymmetric divisors $D_a$ of $X$ are mapped by
$\mu $ to the subspaces $(y,v_a)=0$; here $a=1\dots d$ label the divisors ($d$ here, of
course, is unrelated to the spacetime dimension $d$ of our other sections). 
The   
Sasaki-Einstein $Y$ is given by $X|_{r=1}$, and $r=1$ gives $1=b_ib_jG^{ij}=2(b,y)$.  It is also useful to define $X_1\equiv X|_{r\leq 1}$, with  $\mu (X_1)=
\Delta _b\equiv \{ y | (y,v_a)\geq 0, \ \hbox{and}\ (y,b)\leq \half\}$.   The supersymmetric
$2n-3$ dimensional cycles $\Sigma _a$ of $Y$, for $a=1\dots d$, have cone $D_a=C(\Sigma _a)$ which are the divisors of $X$, and $\mu (\Sigma _a)$ is the subspace ${\cal F}_a$ of  $\Delta _b$ with $(y, v^a)=0$.  

The volume of $Y$ and its supersymmetric cycles $\Sigma _a$ are found from considering 
their cones in $X_1$, which are calibrated by the Kahler form $\omega = dy^i\wedge d\phi _i$. 
This gives \eqn\vols{Vol_b(Y)=2n(2\pi )^n Vol(\Delta _b), \quad Vol_b(\Sigma _a)=(2n-2)(2\pi )^{n-1}{1\over |v_a|}
Vol _b({\cal F}_a).} 
As shown in \MSY, $\sum _a {1\over |v_a|}Vol_b({\cal F}_a)(v_a)_i=2nVol (\Delta _b)b_i$, from 
which it follows that these volumes satisfy
$\pi \sum _a Vol(\Sigma _a)=n(n-1)Vol(Y)$.  (This 
ensures that superpotential terms, associated in the geometry with the holomorphic n-form, have $R(W)=2$.)   

The key point \MSY\ is that the full information of the Sasaki-Einstein metric on $Y$ is not needed to determine the volumes \vols; the weaker information of the Reeb vector $b^i$ and the toric data $v_a$ suffice.  

Moreover, the Reeb vector $b_i$ can be determined from the toric data \MSY.  This fits with the fact that the toric data determines the dual quiver gauge theory  (see e.g. 
\FW\ and references cited therein), from  which the superconformal R-charges can be determined.  The Z-minimization method of
\MSY\ for determining the Reeb vector is to start with the $2n-1$ dimensional Einstein-Hilbert action for the metric $g$ on $Y_{2n-1}$:
\eqn\seh{S[g]=\int _Y (R_g +2(n-1)(3-2n))vol(Y),}
including the needed cosmological constant term associated with the added flux.  Though
\seh\ appears to be a functional of the metric, it was shown in \MSY\ that it's actually only a function of only the Reeb vector:
\eqn\sgsb{S[g]=S[b]=4\pi \sum _a Vol _b(\Sigma_a)-4(n-1)^2Vol_b(Y).}
The full information of the metric is not needed, the weeker information of the Reeb vector suffices to evaluate the action.   

As shown in \MSY,  the condition that $b$ be the correct Reeb vector, associated with a Sasaki-Einstein
metric,  is precisely the condition that the action \sgsb\ be extremal:
\eqn\sbiv{{\partial \over \partial b_i}S[b]=0.}
Defining 
\eqn\zis{Z[b]\equiv {1\over 4(n-1)(2\pi )^n}S[b]=(b_1-(n-1))2nVol(\Delta _b),}
the equation \sbiv\ for $i=1$ gives $b_1=n$, which is just the condition that
the holomorphic n-form transforms as appropriate for a $U(1)_R$ symmetry.  Following
\MSY, define
\eqn\tildezis{\widetilde Z[b_2, \dots b_n]=Z|_{b_1=n}=2nVol _b(\Delta )|_{b_1=n}.}
The equations \sbiv\ for $i\neq 1$ 
give,  upon setting $b_1=n$, 
\eqn\sbivi{0={\partial \over \partial b_i}\widetilde Z[b]=-2(n+1) \int _{\Delta  _b}y_i dy_1\dots dy_n \qquad \hbox{for $i\neq 1$}.}
These are the equations that determine the components $b_i$, for $i=2\dots n$,  of the Reeb vector,
i.e. that pick out the superconformal $U(1)_R$ from the $U(1)^n$ isometry group \MSY. 
The correct Reeb vector {\it minimizes} $\widetilde Z$, since 
the matrix of second derivatives is positive \MSY  
\eqn\sbivv{{\partial ^2\widetilde Z\over \partial b_i\partial b_j}\propto \int _Hy_iy_j d\sigma >0.}

\newsec{$Z$-minimization $=$ $\tau _{RR}$ minimization.}

Let's write \tildezis\ and \vols\ as
\eqn\tildezv{\tilde Z[b_2, \dots b_n]=2nVol_b(\Delta)={1\over (2\pi )^n}Vol_b(Y)|_{b_1=n},}
so $Z$ minimization corresponds to minimizing the volume of $Y$, over the choices of $b_2,\dots ,b_n$, subject to $b_1=n$.  This can be directly related to $\tau _{RR}$ minimization \taumin,
i.e. minimization of the $U(1)_R$ graviphoton's coupling, since  
\eqn\grrsum{\tau _{RR}=C_n {L^{d-3}m^{-(2n+1)}\over 16\pi G_{D_t}}Vol(Y).}
The constant $C_n$ is obtained from adding the contributions \grrccgen\ and \grrkkgen\
and using the relation \tauads.  Let us now consider the quantity 
\grrsum, but with $Vol(Y)$ promoted to the 
function $Vol_b(Y)$, depending on components $b_2, \dots b_n$ 
of the Reeb vector:
\eqn\tauzr{\widetilde \tau _{R_tR_t}[b_2, \dots , b_n]\equiv C_n{L^{d-3}m^{-(2n+1)}\over 16\pi G_{D_t}}Vol_b(Y)=C_n(2\pi )^n{L^{d-3}m^{-(2n+1)}\over 16\pi G_{D_t}}\cdot 
\tilde Z[b_2, \dots , b_n].}
For the superconformal $U(1)_R$ values of $b_2, \dots b_n$, $\widetilde \tau _{R_tR_t}=\tau _{RR}$. 

If we hold $L^{d-3}m^{-(2n+1)}/G_{D_t}$ fixed, \tauzr\ suggests a direct relation between $Z$ and $\tau _{RR}$ minimization.   Physically, we should hold the number of flux units $N$ fixed, i.e. use the flux quantization relation to eliminate  $L^{d-3}m^{-(2n+1)}/G_{D_t}$
in favor of $N/Vol(Y)$.  In particular, for IIB on $AdS_5\times Y_5$
and M theory on $AdS_4\times Y_4$,
\eqn\cnmreln{\eqalign{AdS_5\times Y_5: \ C_n{L^{d-3}m^{-(2n+1)}\over 16\pi G_{D_t}}&={4\pi ^3\over 3}\left({N\over Vol(Y)}\right)^2, \cr AdS_4\times Y_7: \ C_n{L^{d-3}m^{-(2n+1)}\over 16\pi G_{D_t}}&={4\pi ^2\over 3\sqrt{6}}\left({N\over Vol(Y)}\right)^{3/2}.}}
Using these in \grrsum\ shows that, for fixed $N$,  $\tau _{RR}$ is actually {\it inversely} related to $Vol(Y)$.  From
that perspective, it would seem that $Z$ minimization instead {\it maximizes} $\tau _{RR}$,
which is opposite to the result of \taumin\ that the exact superconformal $U(1)_R$ 
minimizes $\tau _{RR}$.   To avoid this, we do not promote the constant
$Vol(Y)$  in the flux relations \cnmreln\ to the function $Vol_b(Y)$ of the Reeb vector, but instead there hold it fixed to its true, physical value.  
Then the function $\widetilde \tau _{R_tR_t}[b]$ \tauzr\ is simply a constant times the function
$\widetilde Z[b]$ of \MSY.

To use the formulae of our earlier sections, consider the Killing vectors  
\eqn\chiis{\chi = \chi _i{\partial\over\partial\phi_i}}
for the $U(1)^n$ isometries of toric $Y_{2n-1}$.  R-symmetries, and in 
particular the Reeb vector, have
$\chi _1=n$, and non-R isometries have $\chi _1=0$.  
As we discussed in sections 5 and 6, the isometry 
$d\phi _\chi \rightarrow d\phi _\chi + A_\chi$ has
an associated $2n-3$ form, which is found from the associated shift  $e^\psi \rightarrow e^\psi + h_\chi (Y)A_\chi $.  For the R-symmetry, this comes from the shift of $d\psi '$, and for
non-R flavor symmetries the shift is via $h_\chi = i_\chi \sigma$.  Using the second equality 
in \hiis, we have 
\eqn\hchiis{h_\chi (Y)=F^{ij}b_i \chi _j =G^{ij}b_i\chi_j =2y^i\chi_i
=2\langle r^2\theta,\chi\rangle ,}
with the inner product with $r^2 \theta$ as in \MSY. For the Reeb vector, \hchiis\ gives
$h_K=1$, since the cone $r=1$ has $1=b_ib_jG^{ij}=2(b,y)$ \MSY.    

For the non-R isometries, we can take as our basis of Killing vectors e.g. $\chi ^{(i)}={\partial \over \partial \phi _i}$, so $\chi ^{(i)}_j=\delta _{ij}$, for $i=2\dots n$.  Then \hchiis\ gives simply
\eqn\hchiiss{h_{\chi ^{(i)}}=2y^i.} 
In this basis, where $U(1)_{F_i}$ is associated with Killing vector ${\partial \over \partial \phi _i}$, the $F_i$ charge of a brane wrapped on cycle $\Sigma$ is
\eqn\fsigmatoric{\eqalign{F_i[B_\Sigma]&=-(n-1)\mu _{2n-3}m^{-(2n-2)}\int _\Sigma 2y^i vol(\Sigma)\cr
&=-n(n-1)\cdot R[B_\Sigma]\cdot {\int _\Sigma y_i vol(\Sigma)\over \int _\Sigma vol(\Sigma)}}.}
In particular, for IIB background $AdS_5\times Y_5$, we have
\eqn\fsigmatoricv{F_i[B_\Sigma]=-{2\pi N\over Vol(Y _5)}\int _{\Sigma  _3}y_i vol(\Sigma ),}
and for M theory background $AdS_4\times Y_7$ we have
\eqn\fsigmatoricvii{F_i[B_\Sigma]=-{4\pi ^2 N\over Vol(Y _7)}\int _{\Sigma _5}y_i vol(\Sigma ).}

Using our formulae from sect. 7, we can determine the kinetic terms
$g^{-2}_{IJ}$, and hence $\tau _{IJ}$ in terms of the geometry of $Y$.  In particular,  
using \grfccgen\ and \hchiiss, we have
\eqn\taurftor{\tau _{RF_i}=C_n{L^{d-3}m^{-(2n+1)}\over 16\pi G_{D_t}}\left(-n(n-1)\right)\int _Y y^i vol(Y),}
with $C_n$ the same constant appearing in \grrsum.  Note that 
\eqn\yiintx{\int _Yy^ivol(Y)=2(n+1)\int _{X_1}y^i vol(X_1)=2(n+1)(2\pi )^n\int _{\Delta _b}y^i dy^1\dots dy^n,}
($2(n+1)$ accounts for the extra $r$ integral in $X_1$). 
Moreover, eqn. (3.21) of \MSY\ gives
\eqn\msyrelnx{\int _{\Delta _b}y^i dy^1\dots dy^n=-{1\over 2(n+1)}{\partial \over \partial b_i}Vol_b(\Delta).}
So \taurftor\ gives
\eqn\taurftorc{\tau _{RF_i}=C_n{L^{d-3}m^{-(2n+1)}\over 16\pi G_{D_t}}(2\pi )^n{(n-1)\over 2}{\partial \over \partial b_i}\widetilde Z[b_2, \dots b_n].}
As discussed, we take the factors in \cnmreln\ to be $b_i$ independent constants, so 
\taurftorc\ can be written as 
\eqn\taurfder{\tau _{RF_i}={(n-1)\over 2}{\partial \over \partial b_i}\widetilde \tau _{R_tR_t}[b_2\dots b_n].}
The relation \taurftorc\ shows that the 
 $\tau _{R_tR_t}$ minimization equations, $\tau _{RF_i}=0$, are  indeed equivalent to
the $Z$ minimization equations  \sbivi\ of \MSY.

We can similarly use our formula \grfccgen\ and \hchiis\ to obtain the coefficient 
$\tau _{F_iF_j}$ for two flavor currents:
\eqn\taufftor{\tau _{F_iF_j}= C_n{L^{d-3}m^{-(2n+1)}\over 16\pi G_{D_t}}\left(n(n-1) \right)^2\int _Y y^i y^jvol(Y),}
with $C_n$ the same constant appearing in \grrsum.   Note now that 
\eqn\yyiintx{\int _Yy^iy^jvol(Y)=2(n+2)\int _{X_1}y^iy^j vol(X_1)=2(n+2)(2\pi )^n\int _{\Delta _b}y^i y^jdy^1\dots dy^n.}
Moreover, in analogy with the derivation of \msyrelnx,  in eqn. (3.21) of \MSY, we find:
\eqn\msyyrelnx{\int _{\Delta _b}y^i y^jdy^1\dots dy^n={1\over 4(n+1)(n+2)}{\partial ^2\over \partial b_i\partial b_j}Vol_b(\Delta).}
We can then write \taufftor\ as 
\eqn\tauffder{\tau _{F_iF_j}={n(n-1)^2\over 4(n+1)}{\partial ^2\over \partial b_i\partial b_j}\widetilde \tau _{R_tR_t}[b_2\dots b_n],}
where again we take  \cnmreln\  as $b$ independent.

Since $\widetilde \tau _{R_tR_t}$ is proportional to $\widetilde Z$,  \tauffder\ provides a way to evaluate the current two-point function coefficients $\tau _{F_iF_j}$ entirely in terms of the Reeb vector and the toric data, without needing to know the metric.  

In \taumin, we discussed the trial function $\tau _{R_tR_t}(s_i)$, which is
quadratic in the parameters $s_i$, and satisfies
\eqn\tauminex{\tau _{R_tR_t}|_{s^*}=\tau _{RR}, \quad {\partial \over \partial s_i}\tau _{R_tR_t}|_{s^*}=2\tau _{Ri}=0, \quad {\partial ^2\over \partial s_i\partial s_j}\tau _{R_tR_t}(s)=2\tau _{ij}.}
This can be compared with the function $\widetilde \tau _{R_tR_t}(b_i)$ defined above, 
which coincides with $\tau _{RR}$ for the minimizing values $b_i^*$, which 
are determined by setting the derivatives to zero, \taurfder, and the second
derivatives \tauffder\ are proportional to $\tau _{ij}$, as in \tauminex.  The relation
between $s_i$ and $b_i$ can be chosen 
 to convert
the coefficients in \tauffder\ to equal those of \tauminex. 

Let us now consider further the expression \fsigmatoric, or more explicitly \fsigmatoricv\
and \fsigmatoricvii, for the flavor charges of branes wrapped on cycles.  We would like to
evaluate these for the supersymmetric cycles $\Sigma _a\subset Y$, i.e. to evaluate
\eqn\yiintsigma{\int _{\Sigma_a}y^i vol(\Sigma)}
in terms of the toric data and Reeb vector.  
Note that 
\eqn\sigmaar{\int _{\Sigma _a}y^i vol(\Sigma)=2n\int _{C(\Sigma _a)}y^ivol(C(\Sigma _a))=
2n(2\pi )^{n-1}\int _{{\cal F}_a}y^i d\sigma _a,}
where the $2n$ factor is from the extra $r$ integral in going from $\Sigma _a$ to $C(\Sigma _a)$, and  $d\sigma _a$ is the measure on ${\cal F}_a$, from $\int \delta ((y, v_a))dy^1\dots dy^n$.   In analogy with the derivation of eqn. (3.21) in
\MSY, it seems likely that the $y^i$ in \yiintsigma\  and \sigmaar\ can be obtained from 
the volume $Vol_b(\Sigma _a)$ in \vols\ by differentiating w.r.t. $b_i$.
 But completing this argument, accounting for all the potential new boundary
terms, seems potentially subtle (to us).   

Let us, instead, note a different way to compute
the charges from the toric data.  Consider the expression for $Vol_b(Y)$,
as a function of both $b$ {\it and} the toric data $(v_a)_i$.  In the integral leading to $Vol_b(Y)=2n(2\pi )^nVol(\Delta _b)$ in \vols, the vectors $(v_a)^i$ appear via
the boundary of $\Delta _b$, which has $(y,v_a)\geq 0$.   Thinking of them 
as variables, taking the derivative w.r.t. $v_a$
then gives a contribution only on the boundary $(y, v_a)=0$:
\eqn\vaider{{\partial \over \partial (v_a)_i}Vol (\Delta _b)=-\int _{{\cal F}_a}y_i d\sigma _a.}
Using \sigmaar\ and \vols\ then gives
\eqn\yiaint{\int _{\Sigma _a}y^i vol(\Sigma)=-{1\over 2\pi}{\partial \over \partial (v_a)_i}Vol_b(Y).}

In the above expressions for $\widetilde \tau _{RR}$ and $\tau _{RF_i}$ and
$\tau _{F_iF_j}$, the Ramond-Ramond and Kaluza-Klein contributions to $g^{-2}_{IJ}$
were summed together, in the coefficient $C_n$. Using the relation \rrktis, which here
gives $(g^{-2}_{IJ})^{CC}=(n-1)(g^{-2}_{IJ})^{KK}$, those two contributions have a fixed
ratio.  Let us now examine that relation in the present context.  For general Killing vectors 
$\chi ^{(I)}$ and $\chi ^{(J)}$, the contribution \gccis\ to their 
mixed kinetic 
term is 
\eqn\gccsy{(g^{-2}_{IJ})^{CC}\propto \int _Y 4y^iy^j \chi _i^{(I)}\chi _j^{(J)}vol(Y).}
The contribution \taukk\ of the Einstein term is similarly 
\eqn\gkksy{(g^{-2}_{IJ})^{KK}\propto \int _Y G^{ij}\chi _I^{(I)}\chi _j^{(J)}vol(Y).}
Taking both $I$ and $J$ to be the R-symmetry, with $\chi _I$ and $\chi _J$ the
Reeb vector, the relation from $(g^{-2}_{IJ})^{CC}=(n-1)(g^{-2}_{IJ})^{KK}$ is 
\eqn\reebkkcc{\int _Y G^{ij}b_ib_j dy_1\dots dy_n= 4\int _Y (y^ib_i)^2dy_1\dots dy_n; }
which is clearly satisfied, since $2b_iy^i=G^{ij}b_ib_j=1$.  For non-R flavor symmetries, the
identity is less trivial.   For general $Y_{2n-1}$ it states that 
\eqn\integralrelation{\int_{Y_{2n-1}} G^{ij} vol(Y)= 4(n-1)^2\int_{Y_{2n-1}}y^iy^j vol(Y)\qquad i,j\neq 1.} 
The extra factor of $(n-1)^2$, as compared with \reebkkcc, is as in \grfccgen, coming from 
writing the volume
form as $\sim e^\psi \wedge (de^\psi )^{n-1}$ and the fact that $\omega _R$ is found from
the shift of the first $e^\psi$ factor, whereas the non-R isometries are 
obtained by shifting the $n-1$ factors of
$d(e^\psi)$.
The relation \integralrelation\ can indeed be verified to hold in the various examples.  
It can also be written in terms of integrals over $\Delta _b$, by extending to $X_1$
and doing the extra $r$ integrals, as 
\eqn\integralrelationx{(n+1)\int _{\Delta _b}G^{ij}dy^1\dots dy^n=4(n-1)^2(n+2)\int _{\Delta _b}y^i y^j dy^1\dots dy^n.}

\newsec{Examples and checks of AdS/CFT: $Y^{p,q}$}

The metric of \refs{\GMSW, \MS} is simply written in the basis of unit one-forms
\eqn\ypqof{\eqalign{e^\psi = {1\over 3}(d\psi '-\cos \theta d\phi + y(d\beta +\cos \theta d\phi))\cr
e^{\theta }=\sqrt{{1-y\over 6}}d\theta , \qquad e^\phi =\sqrt{{1-y\over 6}}\sin \theta  d\phi, \cr
e^y = {1\over \sqrt{wv}}dy, \qquad 
e^{\beta }={\sqrt{wv}\over 6}(d\beta +\cos \theta d\phi ),}}
as $ds^2_Y=(e^\theta )^2+(e^\phi )^2+(e^y)^2+(e^\beta )^2+(e^\psi )^2$. The
coordinate $y$ lives in the range $y_1\leq y \leq y_2$, where $y_1$ and $y_2$ are the
two smaller roots of $v(y)=0$ \MS :
\eqn\yonetwo{y_1={1\over 4p}\left(2p-3q-\sqrt{4p^2-3q^2}\right), \qquad y_2={1\over 4p}\left(2p+3q-\sqrt{4p^2-3q^2}\right).}
The local Kahler form of the 4d base is
\eqn\jypqis{J=e^\theta \wedge e^\phi + e^y \wedge e^\beta.}

The gauge symmetries in $AdS_5$ of IIB on $Y_{p,q}$, and the global
symmetries of the dual SCFTs \BF, are $U(1)_R\times SU(2)\times U(1)_F\times U(1)_B$.
The first three factors are associated with isometries of the metric, and $U(1)_B$
comes from the single representative of $H_3(Y_{p,q},Z)$ (topologically, all
are $S^2\times S^3$).  As usual, the superconformal $U(1)_R$ symmetry is associated with the shift in
$e^\psi$: ${1\over 3}d\psi ' \rightarrow {1\over 3}d\psi ' +{2\over 3}A_R$, and the 
associated 3-form is that of \BHK: 
\eqn\ypqomr{\widehat \omega _R\equiv {1\over 4L^4}\omega _R={1\over 6} e^\psi \wedge J.}
The $SU(2)$ is symmetry is an non-R isometry, associated
with rotations of the spherical coordinates $\theta$ and $\phi$.  Finally, the $U(1)_F$
isometry is associated with shifts $d\beta +\cos \theta d\phi \rightarrow d\beta +\cos \theta d\phi + A_{F}$.  $U(1)_\phi \subset SU(2)$ and $U(1)_F$ form a basis for the $U(1)^2$
non-R isometries, expected from the fact that $Y_{p,q}$ is toric \MS. 
The 3-forms associated with these flavor $U(1)^2$ are found from
\hiis\ and \omegafis\ to be
\eqn\ypqomf{\widehat \omega _\phi \equiv {1\over 4L^4}\omega _\phi = -\cos \theta \widehat \omega _R \qquad \hbox{and}\qquad \widehat \omega _F\equiv {1\over 4L^4}\omega _F=-y \widehat \omega _R.}

The 3-form associated with the $U(1)_B$ baryonic symmetry was already constructed in 
 \HEK, restricting their form $\Omega _{2,1}$ on $C(Y_{p,q})$ to $Y_{p,q}$ by setting
$r=1$:
\eqn\ypqomb{\mu _3\omega _B=  {9\over 8\pi ^2}(p^2-q^2) e^\psi \wedge \eta \qquad \eta \equiv {1\over (1-y)^2}(e^\theta  \wedge e^\phi - e^y \wedge e^\beta),}
where the normalization constant is to keep the periods of $\mu _3 \int C_4$ properly 
integral.

D3 branes wrapped on the various supersymmetric 3-cycles $\Sigma _a$ 
of $Y$ map to the di-baryons of the dual gauge theory \BF\ as:
\eqn\dibarmap{\Sigma _1\leftrightarrow \det Y, \quad \Sigma _2\leftrightarrow \det Z, 
\quad \Sigma _3\leftrightarrow \det U_\alpha, \quad \Sigma _4\leftrightarrow 
\det V_\alpha.}
The cycles $\Sigma _1$ and $\Sigma _2$ are given by the coordinates at  $y=y_1$ and $y=y_2$ respectively \MS.  The cycle 
$\Sigma _3$ is given by fixing $\theta $ and $\phi$ to constant
values, which yields the $SU(2)$ collective coordinate of the di-baryon \HEK. 
The cycle $\Sigma _4\cong \Sigma _2+\Sigma _3$.

As in \BHK, the R-charges of the wrapped D-3 branes, computed from $\mu _3\int _{\Sigma _i }\omega _R$, are
\eqn\ypqrare{R(\Sigma _i)={\pi N\over 3Vol(Y_5)}\int _{\Sigma _i}vol(\Sigma)={\pi N\over 3}{Vol(\Sigma _i)\over Vol(Y_5)}.}  It was
verified in  \refs{\MS, \BBC, \BF, \HEK} that the R-charges computed from the cycle volumes as
in \ypqrare\ agree perfectly with the map \dibarmap\ and the superconformal R-charges, computed in the field theory dual by using the a-maximization \IW\ method.

We can similarly verify that integrating the $U(1)_\phi$, $U(1)_F$ and $U(1)_B$ 3-forms \ypqomf\ and \ypqomb\ over the 3-cycles $\Sigma _a$ agree with the map \dibarmap\ 
and the corresponding charges of the dual field theory \BF. 
 For $U(1)_B$ we have
\eqn\ombypqi{B(\Sigma _i)= \mu _3\int _{\Sigma _i}\omega _B= {9\over 8\pi ^2}(p^2-q^2)\int _{\Sigma _i} 
e^\psi \wedge {1\over (1-y)^2} (e^\theta \wedge e^\phi - e^y \wedge e^\beta),}
and, as already computed in \HEK, this gives (reversing $\Sigma _1$'s orientation)
\eqn\bariypq{B(\Sigma _1)=(p-q), \qquad B(\Sigma _2)=(p+q), \qquad B(\Sigma _3)=p,}
in agreement with the $U(1)_B$ charges of \BF\ for $Y$, $Z$, and $U_\alpha$, respectively.
One minor difference is that we normalize the $U(1)_B$ charges for the bi-fundamentals 
with a factor of $1/N$, so that the charges of the baryons are $O(1)$ rather than $O(N)$; this
is natural when $U(1)_B$ is thought of as an overall $U(1)$ factor of a $U(N)$ gauge group, and
also natural in terms of having the charges be properly quantized, so that $\int \mu _3 C_4$ and $\int B(Q_i)A_B$ are gauge invariant mod $2\pi$ under large gauge transformations. 

We can compute the $U(1)_F$ charges of the wrapped D3 branes by using \fsigmagain, here
with $h=y/3$:
\eqn\fypqint{F(B_\Sigma)=-R(B_\Sigma){\int _\Sigma yvol(\Sigma)\over \int _\Sigma vol(\Sigma)}.}  
This gives
\eqn\fypqis{F(\Sigma _1)= y_1R(\Sigma _1), \qquad F(\Sigma _2)=-y_2R(\Sigma _2), \qquad F(\Sigma _3)=-\half (y_1+y_2)R(\Sigma _3).}
The $\Sigma _1$ and $\Sigma _2$ cases follow immediately from \fypqint,
since $y=y_1$ and $y=y_2$ is constant (the $\Sigma _1$ integral gets an extra minus
sign from the orientation), and $F(\Sigma _3)$ in \fypqis\ simply comes from ${\int _{y_1}^{y_2}ydy/\int _{y_1}^{y_2}dy}$. The charges \fypqis\ agree with the $U(1)_F$
charges of \BF, up to the ambiguity that we have mentioned for redefining $U(1)_F$ by
an arbitrary addition of $U(1)_B$, i.e. $U(1)_F^{here}=U(1)_F^{there}+\alpha U(1)_B$.

Using the metric \refs{\GMSW, \MS}, we can explicitly compute the contributions $\tau ^{CC}_{IJ}$
in \tauadsv\ and the contributions $\tau ^{KK}_{IJ}$ in \tauadsvkk, and verify that
$\tau _{IJ}^{CC}=2\tau _{IJ}^{KK}$, as expected from \taukkcc, for the $U(1)_R$ and
$U(1)_\phi$ and $U(1)_F$ isometry gauge fields.  For $U(1)_B$, there is only
the $\tau _{IJ}^{CC}$ contribution to $\tau _{IJ}$.  
For the superconformal $U(1)_R$, we find, as expected $\tau^{KK}_{RR}=4N^2\pi^3/ 9 Vol(Y_{p,q})$ and
$\tau^{CC}_{RR}=8N^2\pi^3/9Vol(Y_{p,q})$, with \MS
\eqn\volypq{Vol(Y_{p,q})={q^2[2p+(4p^2-3q^2)^{1/2}]\over 3p^2[3q^2-2p^2+p(4p^2-3q^2)^{1/2}]}\pi ^3.}

For $\tau _{FF}^{KK}$, the metric  \refs{\GMSW, \MS}\ gives
$g_{ab}K_F^aK_F^b={1\over 36}wq+{1\over 9}y^2={1\over 36}w(y)$, so \tauadsvkk\ yields
\eqn\tauKKYpqF{\tau^{KK}_{FF}={N^2\pi^3\over36Vol(X_5)}{\int dy
w(y)(1-y)\over\int dy(1-y)}
={N^2\pi^3\over18Vol(X_5)}{\sqrt{4p^2-3q^2}\over
p^2}\left(2p-\sqrt{4p^2-3q^2}\right).}
Using $\widehat \omega _F$ of \ypqomf\ in \tauadsv\ we can also compute  
\eqn\omegaFYpq{\tau^{CC}_{FF}=\tau^{CC}_{RR}{\int dy
y^2(1-y)\over\int dy(1-y)}=\tau^{CC}_{RR}{1\over16}{\int
dyw(y)(1-y)\over\int dy(1-y)}=2\tau^{KK}_{FF},}
satisfying the relation \taukkcc.  
Combining \tauKKYpqF\ and \omegaFYpq\ gives 
\eqn\YpqFres{\tau_{FF}={N^2\pi^3\over6 Vol(Y_{p,q})}{\sqrt{4p^2-3q^2}\over
p^2}\left(2p-\sqrt{4p^2-3q^2}\right).} 
This result for $\tau _{FF}$ can be compared with the field theory prediction.  The
$U(1)_F$ charges of the bifundamentals are found from the $U(1)_F$ charges 
\fypqis\ of the dibaryons, and the map \dibarmap (so the factor of $N$ from
\ypqrare\ is eliminated), e.g. $F(Z)=-y_2R(Z)=-y_2\pi Vol(\Sigma _2)/3Vol(Y_5)$, which looks
rather ugly when written out in terms of $p$ and $q$.  From these charges and the 
$U(1)_R$ charges, we can compute the 't Hooft anomalies, and thereby compute
$\tau _{FF}$ on the field theory side by using the relation $\tau _{FF}=-3\Tr RFF$.  The result
is found to agree perfectly with \YpqFres.  

Let us now consider $\tau _{RF}$.  The Kaluza-Klein contribution is given as in \tauadsvkk,
with $g_{ab}K_R^aK_F^b=y/9$, and the integral over $y$ vanishes, so $\tau ^{KK}_{RF}=0$.  Likewise, $\tau ^{CC}_{RF}=0$, because $\int y(1-y)$ vanishes.  So, as expected, $\tau _{RF}=0$.  

As we discussed in the previous section, the $F_i[\Sigma _a]$ charges and $\tau _{IJ}$ can
also be computed entirely from the toric data and Z-function of \MSY.  In
the toric basis of \MSY,
\eqn\msyvbasis{v_1=(1,0,0),\quad v_2=(1,p-q-1,p-q),\quad
v_3=(1,p,p),\quad v_4=(1,1,0).}
The Z-function is, with $(b_1, b_2, b_3)\equiv (x,y,t)$, \MSY
\eqn\msyzbis{Z[x,y,t]={(x-2)p(p(p-q)x+q(p-q)y+q(2-p+q)t)\over 2t(px-py+(p-1)t)((p-q)y+(1-p+q)t)(px+qy-(q+1)t)},}
which, imposing $x=1$, is minimized for \MSY:
\eqn\bminis{b_{min}=\left(3,{1\over2}(3p-3q+\ell^{-1}),{1\over2}(3p-3q+\ell^{-1})\right),\quad
\ell^{-1}={1\over q}\left(3q^2-2p^2+p\sqrt{4p^2-3q^2}\right).}

Our formula \tauffder, for example, gives $\tau _{F_iF_j}$, for the $F_i$ associated with the
$ \sim {\partial \over \partial \phi _i}$ Killing vectors, in terms of the Hessian of second derivatives of the function \msyzbis, evaluated at \bminis.  To connect the results 
in the toric basis for the flavor symmetries to those discussed above, we note that 
the Killing vector for shifting $\beta$ can be related to those for shifting $\phi _1$
and $\phi _2$ as ${\partial\over\partial\beta}={\ell^{-1}\over6}\left({\partial \over \partial \phi _2} +{\partial \over \partial \phi _3}\right)$, so $U(1)_F={\ell^{-1}\over6}(U(1)_2+U(1)_3)$.

\centerline{\bf Acknowledgments}

We would like to thank Mark Gross, Dario Martelli, Dave Morrison,  Ronen Plesser, and Nick Warner, for discussions. KI thanks the ICTP Trieste and CERN for hospitality during
the final stage of writing up this work, and the groups and visitors there for discussions.   This work was supported by DOE-FG03-97ER40546.

\listrefs

\end